\documentclass[iop]{emulateapj}
\usepackage{amssymb}
\usepackage{amsmath}
\usepackage{graphicx}
\usepackage[normalem]{ulem}
\usepackage{color}
\newcommand{\colj}[1]{{\textcolor{black}{#1}}}
\newcommand{\colb}[1]{{\textcolor{black}{#1}}}

\newcommand{\vvnew}[1]{{\textcolor{black}{#1}}}
\newcommand{\s}{$\,$}
\newcommand{\vnew}[1]{{\textcolor{black}{#1}}}
\newcommand{\postsub}[1]{{\textcolor{black}{#1}}}

\tabletypesize{\scriptsize}
\newcommand{\targ}{2M2139~}
\newcommand{\targs}{2M2139}

\shorttitle{Variability of an L/T Transition Brown Dwarf}
\shortauthors{Radigan et al.}

\begin{document}

\title{Large Amplitude Variations of an L/T Transition Brown Dwarf:  Multi-Wavelength Observations of Patchy, High-Contrast Cloud Features}

\author{Jacqueline Radigan$^1$, Ray Jayawardhana$^1$, David Lafreni{\`e}re$^2$, {\'E}tienne Artigau$^2$, Mark Marley$^3$ and  Didier Saumon$^4$}
\affil{$^1$ Department of Astronomy, University of Toronto, 50 St. George Street, Room 101, Toronto, Canada M5S 3H4 $^2$ D{\'e}partement de Physique and Observatoire du Mont M{\'e}gantic, Universit{\'e} de Montr{\'e}al, Montr{\'e}al, Canada, $^3$NASA Ames Research Center, Moffett Field, CA, United States, 94035, $^4$Los Alamos National Laboratory, Los Alamos, NM 87545}



\slugcomment{Submitted to the Astrophysical Journal}

\begin{abstract}
We present multiple-epoch photometric monitoring in the $J$, $H$, and $K_s$ bands of the T1.5 dwarf 2MASS J21392676+0220226 (2M2139), revealing persistent, periodic \postsub{($P=7.721\pm$0.005\s hr)} variability with a peak-to-peak amplitude as high as 26\% in the $J$-band. The light curve shape varies on a timescale of days, suggesting that evolving atmospheric cloud features are responsible.  Using interpolations \vnew{between model atmospheres with differing cloud thicknesses} to represent a heterogeneous surface, we find that the multi-wavelength variations and the near-infrared spectrum of \targ can be reproduced by either (1)cool, thick cloud features sitting above a thinner cloud layer, or (2)warm regions of low condensate opacity in an otherwise cloudy atmosphere, possibly indicating the presence of holes or breaks in the cloud layer.  \vnew{We find that temperature contrasts between thick and thin cloud patches must be greater than $175$\s K and as high as 425\s K.}
We also consider whether the observed variability could arise from an interacting binary system, but this scenario is ruled out.  \targ joins the T2.5 dwarf SIMP0136 discovered by Artigau and coworkers as the second L/T transition brown dwarf to display large-amplitude variability on rotational timescales, suggesting that the fragmentation of dust clouds at the L/T transition may contribute to the abrupt decline in condensate opacity and $J$-band brightening observed to occur over this regime.
\end{abstract}


\section{Introduction}
 With observed temperatures ranging from $\sim$2200-300~K, field brown dwarfs (BDs) with L and T spectral types possess the coolest atmospheres currently available to direct and detailed study, and thus constitute important precursors to investigating directly imaged extra-solar giant planet atmospheres.  At these cool temperatures, an understanding of atmospheric condensates---which affect luminosities, radii, cooling histories, and spectral morphologies---is essential to deriving correct physical properties for both cool BDs and giant planets \citep[e.g.][]{marley02,burrows11}.

Our current understanding of ultracool atmospheres, including the formation and sedimentation of condensate clouds, has developed based on comparisons of detailed atmosphere models to observations of hundreds of L and T dwarfs identified in the solar neighborhood \citep[e.g.,][]{cushing08,stephens09,witte11}.  Nonetheless, fundamental questions remain concerning the most basic properties of condensates, including their vertical and horizontal distributions, how these evolve as a function of effective temperature, and the role of secondary parameters such as  gravity, metallicity, convection, and rotation.

Differences in condensate clouds are responsible for a wide range of spectral morphologies at a given effective temperature, as illustrated by the transition between cloudy L and clear T spectral types.  A detailed review of the L and T spectral sequence can be found in \citet{kirkpatrick05} and references therein.  In short, the L spectral sequence is characterized by the formation and thickening of condensates with decreasing effective temperature, the L/T transition ($\sim$L7-T4 spectral types) by a significant decrease in condensate opacity at a roughly constant effective temperature \citep{golimowski04,stephens09}, and the mid-to-late T spectral sequence by the cooling of a cloud-free atmosphere, in which condensates have gravitationally settled below the photosphere. 

Whereas the predominantly cloudy and predominantly clear end-points of the L/T transition have been well-modeled by various groups \citep{chabrier00,allard01,tsuji02,allard03,burrows06,saumon08}, the transition regime 
remains poorly understood.  
While a progressive reddening in NIR $J-K_s$ color is seen throughout most of the L spectral sequence due to an increasing photospheric dust opacity, the L/T transition is evident as an abrupt blue-ward shift in $J-K_s$ color by $\sim$2 magnitudes, accompanied by
a brightening in the $J$-band of up to $\sim$1 mag \citep[e.g.][Faherty et al. 2011, submitted]{vrba04,zapatero07}. This dramatic blue-ward evolution and $J$-band brightening
are not well-reproduced by 1D atmosphere models incorporating condensate formation and sedimentation \citep{ackerman01,allard03,burrows06,saumon08} which predict a much more gradual turn around of $J-K_s$ color with decreasing effective temperature.  
Various ideas have been put forward to explain the L/T transition including a sudden rain-out of condensates (or equivalently an evolving sedimentation efficiency) \citep{knapp04}, or a fragmentation of the cloud layer \citep{ackerman01,burgasser02_lt}.  Both scenarios can reproduce main features of the transition such as $J$-band brightening or the re-emergence of molecular FeH in early T dwarfs, which becomes depleted above the cloud layer due to condensate formation.

However, neither idea has been modeled in detail as a physical process: both simply represent requirements that, if satisfied, could reproduce the observed behavior of BDs on a color-magnitude diagram.  In this respect, the cloud fragmentation hypothesis is attractive due to its ability to make testable predictions:  cloud structures, if possessing large enough azimuthal asymmetries, should give rise to photometric variability on rotational timescales.  Furthermore, in our own Solar System clouds on giant planets form discrete patterns coupled to rotationally-dominated atmospheric flows.  For Jupiter, clearings in the cloud layer are evident as 5\,$\mu$m hot-spots, and may result in rotationally modulated variability as high as 20\% \citep{gelino00}.

The combination of condensate clouds and rapid rotation has long-motivated searches for cloud and weather-related variability of ultracool dwarfs \citep[e.g.][]{tinney99}.  Observations in the red optical  \citep{bailer-jones01,gelino02,koen03,koen04b,koen05b,koen05c} have targeted mainly late-M and early-L dwarfs due to a drop-off in optical flux for later spectral types.  Evidence for periodic variability is found for $\sim$30\% of early-type objects monitored in the $I$-band with peak-to-peak amplitudes of a few percent.  However, it is unclear whether modulation is the result of magnetic spots, dust clouds or a combination of both.  
In order to study variability at the L/T transition,
a move to infrared wavelengths wherein late-L and T dwarfs are bright is required.  

Since the atmospheres of late-L and T dwarfs are increasingly neutral, they are less likely to support cool magnetic spots \citep{gelino02,mohanty02}, making the interpretation of detected variability in this regime less ambiguous.  
However, in contrast to optical surveys, ultracool dwarf variability at near and mid infrared wavelengths \citep[e.g.][]{enoch03,koen04a,koen05a,morales-calderon06,goldman08,bailer-jones08,clarke08} appears to be rare (at least at similar amplitudes).
In a study of 18 L and T dwarfs, \citet{koen04b} found no significant evidence of variability in the $J$ band above the $\sim$20 mmag level nor in the $H$ or $K_s$ bands above $\sim$40 mmag.  A further study focussing on known $I$-band variables \citep{koen05a} failed to detect any variability in the near infrared (NIR) $J$, $H$, and $K_s$ bands in any of these same objects.  
In contrast, \citet{enoch03} monitored 9 objects with L2-T5 spectral types in the $K_s$ band and claim large-amplitude periodic variability (0.1-0.4 mag) for 3 objects in their sample.  However, these detections have not been repeated at later epochs, and their amplitudes are only $\sim$1-3 times the reported photometric precision.  

While detections at the level claimed by \citet{enoch03} have not been reported since, recent results from high-precision NIR monitoring provide some evidence for lower-amplitude variability in the NIR.
In a survey of 8 late-L and T dwarfs in the $J$ band, \citet{clarke08} claim periodic variability for 2 of 7 late-L and T dwarfs monitored with amplitudes of 15 and 8 mmag and periods of 1.4~hr and 2~hr respectively.  Most recently, in a single targeted study, \citet{artigau09} found the T2.5 dwarf, SIMP0136 \citep{artigau06}, to be variable with a peak-to-peak amplitude of $\sim$50~mmag in $J$ and a period of 2.3~hr (a 10-$\sigma$ detection).  The high significance, repeatability and multi-band detections for SIMP0136 represents a breakthrough in comparison to previous work. Notably, SIMP0136 falls directly within the L/T transition regime.

Here we present continuous photometric monitoring in the $J$, $H$, and $K_s$ bands of the T1.5 dwarf 2MASS J21392676+0220226 \citep[][\targ hereafter]{reid08} using the Wide Field Infrared Camera on the DuPont 2.5-m telescope at Las Campanas. \targ is the most variable BD to date, and was discovered in our large $J$-band monitoring program for weather-related variability of cool BDs (Radigan et al., in preparation).  In \S\ref{sect:obs} we describe the observations and data reduction, and in \S\ref{sect:result} we present and analyze the reduced light curves.  In \S\ref{sect:phys} we present an overview of \targ's physical properties based on existing archival data.  In \S\ref{sect:model} we attempt to model the observed multi-band variability using linear combinations of 1-dimensional cloudy and clear atmosphere models.  In \S\ref{sect:discuss} we discuss our results and possible origins of the observed variability.  Finally, in \S\ref{sect:concl} we give a brief summary of our conclusions and suggestions on how to proceed.

\section{Observations and Data Reduction} 
\label{sect:obs}

\subsection{LCO Observations}

\begin{deluxetable*}{lccccccc}
\tablecolumns{8}
\tablewidth{7in}
\tablecaption{Observing Log \label{tab:log}}
\tablehead{ 
  \colhead{Date\tablenotemark{a}} & \colhead{Instr.\tablenotemark{b}} & \colhead{Filter}  & \colhead{UT Start\tablenotemark{c}} & \colhead{Length (hr)} & \colhead{t$_{{\rm exp}}$ (s)} & \colhead{N$_{{\rm exp}}$} & \colhead{Notes}
}
\startdata
01 Aug 2009 & WIRC & $J$ & 00h19m & 3.1& 40 & 140 &  Clear at start with light clouds after 1.5~hr, 0.7\arcsec~seeing\\
21 Sep 2009 & WIRC & $J$ & 00h14m & 5.12 & 60 & 255 &   Clear with occasional passing light cloud, 0.7\arcsec~seeing\\ 
22 Sep 2009 & WIRC & $K_s$ & 23h53m & 5.48 &20 & 412 &   Clear with occasional passing light cloud, 0.8\arcsec~seeing\\
23 Sep 2009 & WIRC & $J$ & 23h29m & 5.99 & 60 & 260 &   Clear with occasional passing light cloud, 0.7\arcsec~seeing\\
24 Sep 2009 & WIRC & $H$ & 23h25m & 6.01 & 20 & 498 &   Clear with occasional passing light cloud, 0.7\arcsec~seeing \\
26 Sep 2009 & WIRC & $J$,$H$,$K_s$ & 23h45m & 5.58 &45,20,20 & 91,130,132 &  Clear, 1.5\arcsec~seeing\\
30 Sep 2009 & WIRC & $J$,$H$,$K_s$ & 23h40m & 4.81 &45,20,20 & 82,110,114 &  Clear at start becoming cloudy after first 4.1 hr, 1.2\arcsec seeing \\
01 Oct 2009 & WIRC & $J$,$H$,$K_s$& 23h46m & 5.11 &45,20,20 & 77,104,110 &  Clear, 1.1 \arcsec~seeing \\
08 Nov 2009 & CPAPIR & $J$ & 22h37m &  4.37 & 21.6 & 387 &  Clear, 1.5-2\arcsec~seeing
\enddata
\tablenotetext{a}{Dates correspond to the local day at the beginning of the night.}
\tablenotetext{b}{WIRC observations were conducted on the LCO 2.5-m telescope, and CPAPIR on the OMM 1.6-m telescope}
\tablenotetext{c}{UT Start time may be $+$1 day ahead of the local date.}
\end{deluxetable*}

Observations of \targ were made using the Wide Field Infrared Camera \citep[WIRC; ][]{persson02} on the Du Pont 2.5~m telescope at Las Campanas, as part of a large survey for $J$-band variability of cool BDs (Radigan et al., in prep). The camera consists of 4 HAWAII-I arrays, each with a 3.2\arcmin~field of view and a pixel scale of 0.2$\arcsec$.  The camera is intended as a wide-field survey camera, with 3\arcmin~gaps between detectors.  We have not used it as such, choosing to position our target consistently on the south-west array, which we determined to be the least noisy of the four chips.
New $J$-band filters closely matching the Mauna Kea Observatory (MKO) system \citep{tokunaga05} were purchased and installed for our survey in order to minimize the effects of differential atmospheric extinction, by cutting off time-variable telluric water absorption bands red-ward of 1.35 $\mu$m \citep[e.g,][]{artigau06t}.

\targ was first monitored in the $J$ band on 01 Aug 2009 for 2.5 hours. 
An analysis of its light curve revealed a 90\s mmag increase in brightness over this time span.  This prompted us to follow-up with longer photometric sequences in $J$, $H$, and $K_s$ bands on the four consecutive nights of 21-24 Sep 2009.  In order to obtain near-simultaneous photometry in multiple bandpasses, additional photometric sequences alternating between $J$, $H$, and $K_s$ bands every $\sim$20\s min were obtained on 26 Sep 2009, 30 Sep 2009, and 01 Oct 2009.  A log of all observations is provided in Table \ref{tab:log}.

All observations except those of 21 Sep 2009 were made using a random dither pattern wherein the telescope was offset by at least 3\arcsec (15 pixels) after each exposure, and all pointings were contained in a 15" (75 pixels) square box.  Individual exposure times in $J$ were 40\s s  (09 Aug 2009), 45\s s (26 Sep 2009, 3 Sep 2009, 01 Oct 2009), or 60\s s (23 Sep 2009).  Exposures were read out using correlated double sampling.  Along with telescope offsets this resulted in effective cadences of 62\s s, 68\s s, and 82\s s respectively for the listed exposure times.  For the $H$- and $K_s$-bands we used individual exposure times of 20\s s throughout, resulting in an effective cadence of 42\s s. 

The $J$-band observations of 21 Sep 2009 employed a staring strategy, in which the target centroid was kept fixed on the same pixel throughout the sequence.  This was accomplished using an $IDL$ routine to stream the incoming science images onto a standard laptop and compute real-time guiding corrections.  An alert was sounded, and manual closed-loop corrections to the guide-camera reference position were made each time the target strayed by more than 0.5 pixels from its initial position (approximately once every 5-15 minutes).  Without telescope offsets the efficiency of staring observations is significantly increased.  Individual exposures of 60\s s were used, resulting in a cadence of 67\s s.  Nine-point dither sequences for the purpose of rough sky subtraction and centroiding were made at the beginning, middle and end of the 5.12\s hr observation.
 
 For each sequence, dome-flats (lamp on and off) and dark frames corresponding to each exposure time were taken either on the afternoon preceding, or the morning following each observation. 

\subsection{Reduction and Processing of WIRC data}
All raw images were corrected for non-linearity using a detector response curve measured on 27 Jul 2009.  Calibration images were median-combined to create high signal-to-noise dark and flat-field frames.  The dark-current contribution was then subtracted from all other images.  For the dithered sequences, a running sky frame was computed by median combining the 11(7) $J(HK)$-band images that were (i)taken closest in time to, and (ii)were spatially offset by at least 6\arcsec~(30 pixels) from the image being reduced.  After an initial first-pass reduction,  stars were identified from the stacked field, and then masked for the second pass so as not to bias the median-combined sky frames. After sky subtraction, pixel-to-pixel variations in quantum efficiency were removed by dividing the resultant image by the flat-field.  Bad pixels identified from the flat-field or pixels having more than 35000 counts ($>$3\% non-linear) were flagged.
Except for sky subtraction, reduction of the non-dithered staring sequence is almost identical to the procedure described above.  In this case the sky frames were constructed by median combining the 9-point dither sequences taken before and after staring, linearly interpolating between them in time, and scaling to the 3$\sigma$-clipped median of each science image.  Since immediate background levels are also subtracted using aperture photometry, the primary motivation of sky subtraction for the staring sequence is to achieve better centroiding of the target and reference PSFs.  For all reductions and analyses we considered only the south-west array of the WIRC camera.

\subsubsection{Relative Differential Photometry}
For each monitoring sequence, aperture photometry was performed on \targ and a set of reference stars in \targs's the field of view, using a circular aperture of 1.5 times the median full width at half maximum (FWHM) of all stars in each image.   A large aperture was chosen to decrease the systematic effects of slightly elongated PSFs on our photometry which occasionally occur due to a gradual degradation of the telescope focus during long sequences.  Residual sky levels in the vicinity of each star were measured inside an annulus centered on each source of inner radius 3.7 times the FWHM and a width of 11 pixels. Flux measurements in which a flagged pixel fell inside the aperture were set to an error value and disregarded in the light curve analysis.  

The raw light curves display fluctuations in brightness due to changing atmospheric transparency, airmass, and residual instrumental effects throughout the night.  To a very good approximation these changes are common to all stars, and can be removed.  First, the raw light curves of all stars were converted from absolute to relative fluxes via division by their median brightness.  Next, for each reference star a calibration curve was created by median combining the light curves of all other reference stars (excluding that of the target and star in question).  The raw light curve of each reference star was then divided by the corresponding calibration curve to obtain a corrected light curve.  \colj{The standard deviations of the corrected light curves for each reference star  ($\sigma$) were then measured.  This process was repeated multiple times, using an iterative approach where only reference stars with $\sigma<15$\s mmag in $J$ and $<$25\s mmag in $H$ and $K_s$ were kept.\footnote{For the observations of 30 Sep 2009 these criteria were relaxed to $\sigma<20$\s mmag in $J$ and $<$30\s mmag in $K_s$, and for the observations of 01 Oct 2009 they were relaxed to $\sigma<$30\s mmag in $K_s$.} This process was terminated (usually after the second iteration) when the number of good calibration stars no longer changed. }

\begin{figure}
\epsscale{1.2}
\plotone{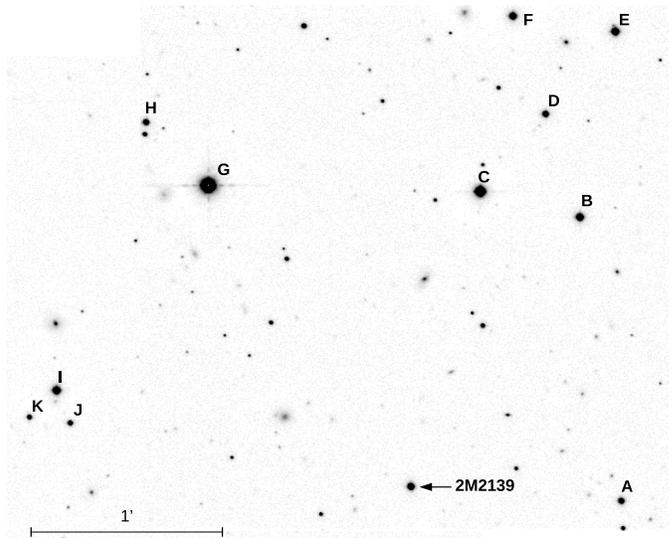}
\caption{Stacked and mosaicked WIRC $J$ band image of the field surrounding \targ.  The target and reference stars used for differential photometry are labeled, with details provided in table  \ref{tab:refs}. \label{fig:fchart}}
\end{figure}

A calibration curve for \targ was then computed using this subset of high signal-to-noise ($S/N$) references, less a comparison star of similar brightness.   Figure \ref{fig:fchart} shows a stacked and mosaicked image of the field with reference stars labelled by letter. \colj{Details (identifiers and magnitudes) of the reference and comparison stars used to calibrate each set of observations are provided in table \ref{tab:refs}}.  We note that on some nights otherwise high $S/N$ reference stars were excluded due to consistently falling on bad pixels, saturation, or contamination from a faint nearby source in poor seeing conditions (e.g. star E).  While it would be optimal to use a common set of reference stars throughout, we find that the shape and amplitude of our final light curves are independent of our choice of reference stars.   For most of our observations star $B$ from figure \ref{fig:fchart} was chosen as the comparison, with the exception of 01 Aug 2009 when it didn't fall in the field of view and we used star $I$ instead.  

\begin{figure*}[float]
\epsscale{0.9}
\plotone{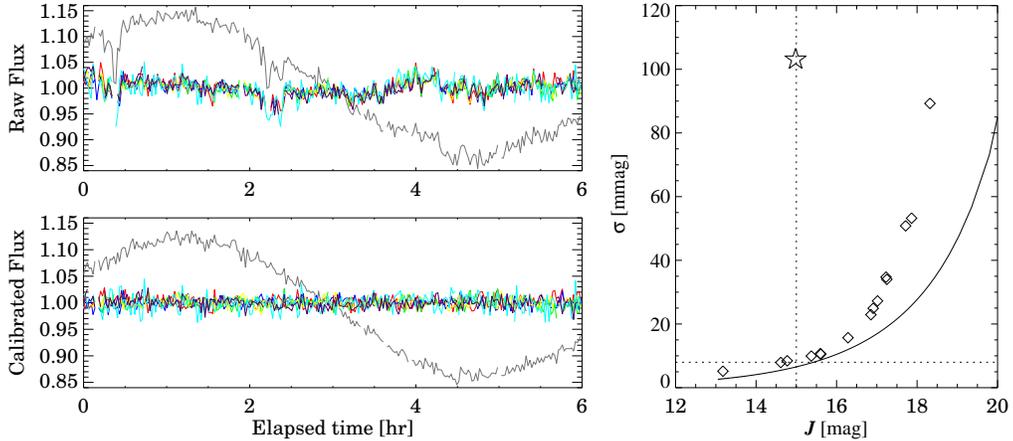}
\caption{\colj{{\em Left:} An example of raw (top) and detrended (bottom) lightcurves for \targ (grey lines) and reference stars (colored lines) for a photometric sequence taken on 23 Sep 2009.   {\em Right: } Standard deviations of the detrended light curves for all stars on the chip as a function of $J$ magnitude.  2M2139 is shown as an open star symbol.  A solid line shows the photon noise limit.  At $J$ magnitudes brighter than $\sim$16.5 observations are close to being photon-limited. \label{fig:diag}}}
\end{figure*}

Detrended light curves for \targ and the comparison star were obtained by dividing their raw fluxes by the final calibration curve.  \colj{An example of raw and detrended light curves for the reference stars and target, as well as light curve standard deviations as a function of star brightness for the 23 Sep 2009 epoch are shown in figure \ref{fig:diag}.  Detrended $J$, $H$ and $K_s$ light curves for \targ and the comparison star for the consecutive nights of 21-24 Sep 2009 are displayed in figure \ref{fig:jhk}.}

\colj{In addition, light curve properties for all epochs including amplitudes, and standard deviations for the target ($\sigma_t$) and comparison star ($\sigma_c$) are provided in table \ref{tab:red}. Since the light curves for \targ show large trends, $\sigma_t$ is obtained by taking the standard deviation of \targ's light curve subtracted by a shifted version (by one element) of itself, and then divided by $\sqrt{2}$.  For the epochs where we cycled through the $J$, $H$, and $K_s$ filters, large time gaps exist in the sequence for any given filter, necessitating a different method for obtaining $\sigma_t$.  Thus for these sequences we took $\sigma_t$ to be the average of the standard deviations obtained for individual $\sim$20\s min segments in a given filter.}

\begin{deluxetable}{lcccc}
\tablecolumns{5}
\tablewidth{3.3in}
\tablecaption{\colj{Reference stars used for differential photometry}\label{tab:refs}}
\tablehead{ 
\colhead{2MASS ID} & \colhead{Letter ID\tablenotemark{a}} & \colhead{$J$} & \colhead{$J$-$K_s$}&\colhead{Epochs used \tablenotemark{b}}
}
\startdata
J21392216+0220185 & A & 15.51 & 0.65 & 2,4,9 \\
J21392311+0222009 & B & 14.46 & 0.64 & 2-9\tablenotemark{c}    \\
J21392542+0222102 & C & 12.98 & 0.48 &  1-3,5-9 \\
J21392392+0222383 & D & 15.22 & 0.93 &   2-5,6-7($J$, $H$ only),8-9\\
J21392228+0223082 & E & 14.26 & 0.64 &   2,4-7   \\
J21392465+0223140 & F & 14.63 & 0.92 &    2-9  \\
J21393173+0222126 & G & 11.48 & 0.37 &     3,5-9 \\
J21393318+0222356 & H & 15.40 & 0.52 &    1  \\
J21393533+0220584 & I & 13.50 & 0.36 &     1\tablenotemark{c} \\
J21393502+0220466 & J & 15.33 & 0.59 &    1  \\
J21393596+0220488 & K & 15.65 & 0.59 &    1  
\enddata
\tablenotetext{a}{Stars are labelled by letter in figure \ref{fig:fchart}.}
\tablenotetext{b}{Epochs from table \ref{tab:log} are numbered chronologically from 1-9.}
\tablenotetext{c}{Comparison star}
\end{deluxetable} 

\subsection{OMM Observations}
Due to the evolving nature of the light curve, the target was observed again on the night of 08 Nov 2009 using the Observatoire du Mont M{\'e}gantic 1.6-m telescope and the Camera PAnoramic Proche Infra-Rouge \citep[CPAPIR,][]{artigau04}. \colj{The camera has a 30\arcmin~field of view with a pixel scale of 0.89\arcsec.  \colj{As with our WIRC observations, an MKO $J$ filter was employed.} Observations commenced at an airmass of 1.39 and concluded at an airmass of 2.35.}  Random dithers were made between each exposure, always keeping the target within 1\arcmin~of the central position.  A single co-addition with an exposure time of 21.6\s s was used for all science frames, resulting in a median cadence of 38\s  s.  

\subsection{Reduction of CPAPIR data}
The raw CPAPIR images were processed using the CPAPIR pipeline.  A running median sky image were constructed by taking the median of the 11 frames centered on the frame of interest after masking bright field stars identified in the Two Micron All Sky Survey \citep[2MASS;][]{2mass} . After sky subtraction, images are divided by a flat field image constructed from on/off dome images. The Poisson noise associated with the flat field is $\sim$0.2\% per pixel and a negligible contributor to the overall error budget.  Aperture photometry was performed in a similar manner to the WIRC images, but using an aperture size of 1 FWHM, and an annulus of inner radius 3.5 FWHM, and 11 pixel width.

\subsection{Photometric Calibration}
In addition to the relative photometry, all light curves were flux-calibrated against the 2MASS catalog using references identified in figure \ref{fig:fchart}.  While the WIRC $H$ and $K_s$ filters are similar to those from 2MASS, the MKO $J$ filter installed for our survey is significantly narrower than the 2MASS $J$ filter.  Because we do not have enough reference stars to derive robust color-dependent corrections between our MKO $J$ filter and 2MASS $J$ filter, we first converted 2MASS $J$ magnitudes of the reference stars to MKO magnitudes according to the color-dependent transformations provided by \citet{leggett06}.  

The 2MASS catalog magnitudes, corrected for our filter system, were then converted to counts (multiplied by some arbitrary factor) and a linear fit between the catalog and measured counts was performed.  The quality of the photometric calibration was determined from the slope error for each linear fit, and found to be accurate among reference stars at the 1-2\% level. 

In figures we have opted to plot light curves in units of relative flux as opposed to magnitudes.  The reference magnitudes corresponding to relative fluxes of 1 are given by $J_{MKO}$=14.75, $H$=14.11, and $K_s$=13.59.  The chosen reference magnitudes correspond to the median magnitudes measured in each bandpass for the near-simultaneous $JHK_s$ sequence of 26 Sep 2009, where the light curves were observed to have similar amplitudes in all bandpasses.

In addition to placing all light curves on a common flux scale, the photometric calibration allows us to compare the magnitudes and colors derived here with archival data from 2MASS and the SpeX Prism Library\footnote{http://pono.ucsd.edu/$\sim$adam/browndwarfs/spexprism/}.  In the SpeX Prism Library there are two low resolution NIR spectra (R$\sim$120) for \targ, obtained in 2003 and 2004 by \citet{burgasser06}.  The SpeX Prism Library, while not able to provide absolute photometry, allows us to derive colors for \targ at additional epochs.  When synthetic 2MASS colors for M, L and T dwarfs from the SpeX Prism Library are compared to their 2MASS values, the match is surprisingly good, with a standard deviation similar to the reported photometric errors from the 2MASS catalog.  For L and T dwarfs no significant systematic trends exist as a function spectral type, whereas we find the synthetic $J-K_s$ colors of $M$ dwarfs to be on average slightly bluer than values derived in 2MASS.  We present a comparison between 2MASS and synthetic SpeX colors in Appendix \ref{sect:ap1}.  

Finally, for a proper comparison between WIRC, 2MASS, and SpeX (synthetic 2MASS) photometry,  we have converted  WIRC magnitudes derived for \targ to the 2MASS system.   Correction terms between the two filter systems were computed directly from \targs's NIR spectrum
using 2MASS and WIRC filter plus system transmission curves and a Kurucz model Vega spectrum as a zero-magnitude flux reference \citep{kurucz79,kurucz93}.  Recall that for reference stars corrections between $J$ filters were accounted for in advance by our conversion of 2MASS to MKO $J$ magnitudes, while $H$ and $K_s$ band corrections are generally negligible.  Magnitudes and colors measured for \targ from WIRC, 2MASS and SpeX epochs, as well as correction terms specific to \targ used to convert between WIRC and 2MASS filter systems, are provided in table \ref{tab:phot}.  

\begin{figure*}[float]
\epsscale{1}
\plotone{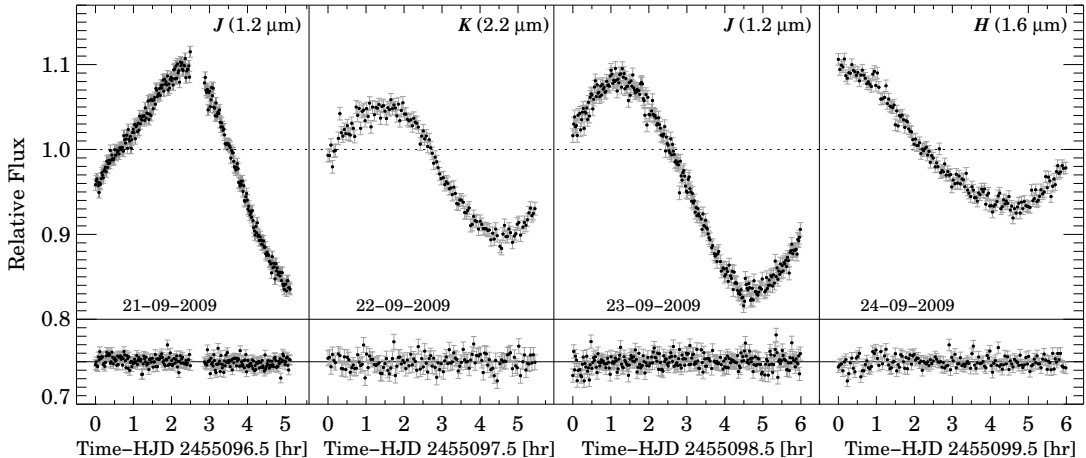}
\caption{Reduced light curves for \targ obtained from 21-24 Sep 2009 (top) and a comparison star of similar brightness (bottom).  Vertical lines separate observations on different nights.  The given dates correspond to the beginning of each observing night. The $J$-band data points show unbinned 60\s s exposures, while the $H$ and $K_s$ data points show 20\s s exposures binned by a factor of 3. \label{fig:jhk}}
\end{figure*} 

\section{Results and Analysis}
 \label{sect:result}
\subsection{The Reduced Light Curves}
Full $J$, $H$, and $K_s$ light curves from the adjacent nights of 21-24 Sep 2009 are shown in figure \ref{fig:jhk}.  \colb{These epochs encompass the largest variability observed for \targ with peak-to-peak amplitudes of  26\% in $J$, 17\% in $H$ and 16\% in $K_s$. }  Peak-to-peak amplitudes are measured directly as the maximum change in brightness, divided by the mid-brightness.  We note that the $H$-band variation may be slightly larger since its global maximum is not entirely captured.  

\begin{figure*}[float]
$\begin{array}{c}
\hspace{1cm}
\includegraphics[width=6in]{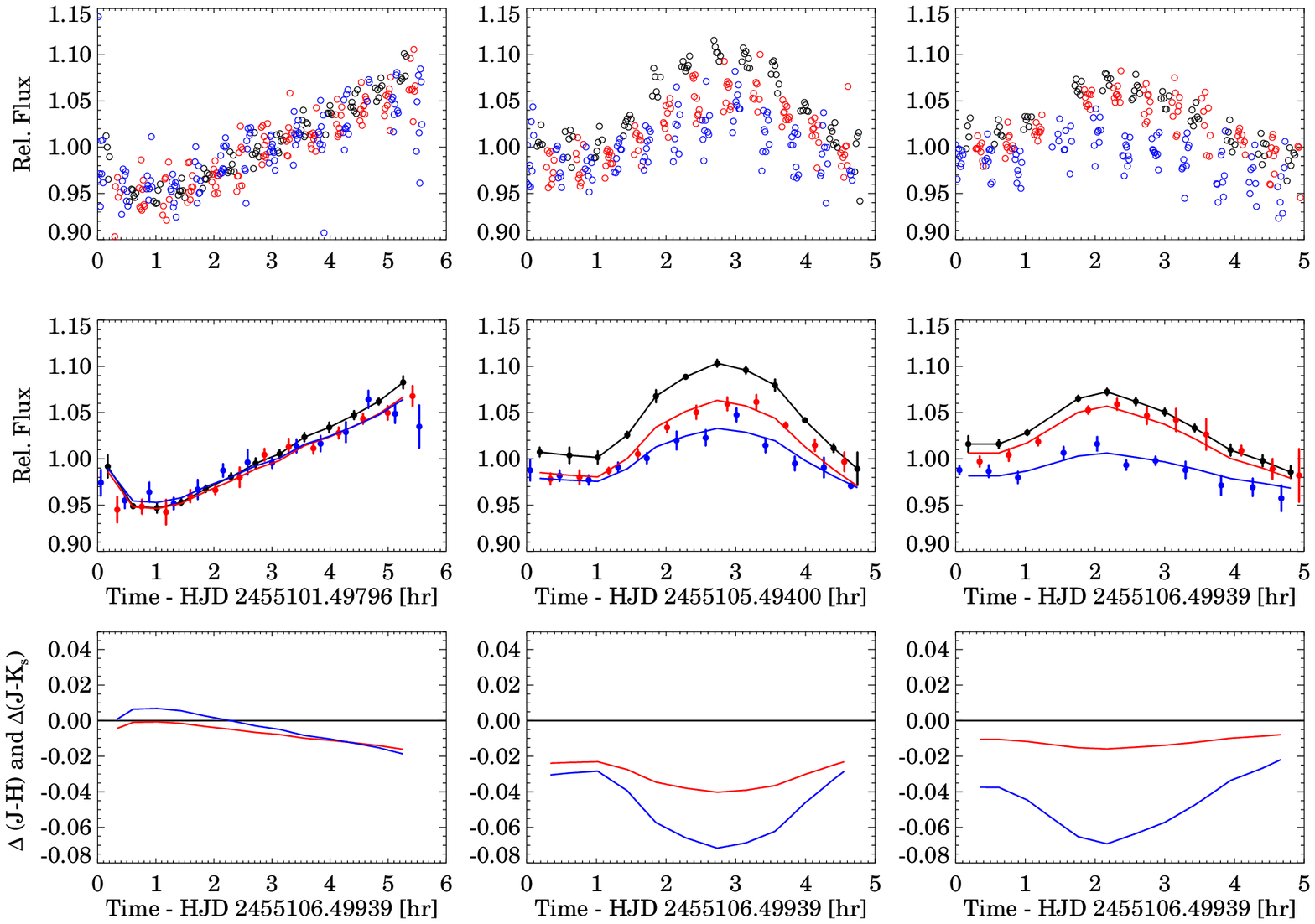} \\
\hspace{1cm}
\includegraphics[width=3.0in]{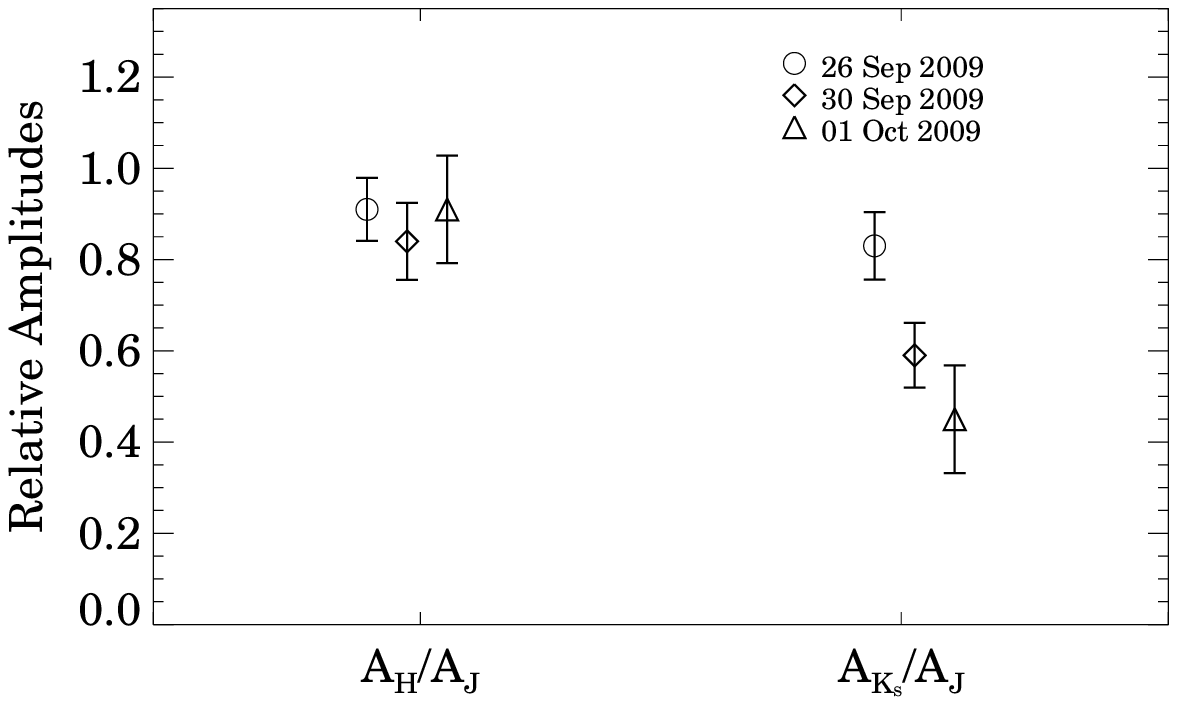}
\end{array}$
\caption{{\em Top row panels}: Unbinned interleaved $J$ (black), $H$ (blue), and $K_s$(red) photometric sequences from 26, 30 Sep 2009 and 01 Oct 2009 epochs.  Unity corresponds to $J_{MKO}$, $H$, and $K_s$ magnitudes of 14.75, 14.11, and 13.59 respectively.  {\em Middle row panels:} Binned (see text for details) $J$, $H$, and $K_s$ light curves, with same color scheme as above.  Linear interpolations of the binned $J$ band points are shown as black lines.  Scaled, best-fitting versions of these $J$-band templates to the $H$ and $K_s$ light curves are shown as red and blue lines respectively.
 {\em Bottom row panels:} Changes in $J-H$ and $J-K_s$ colors, based on the scaled templates.
{\em Bottom panel:} Amplitude ratios and uncertainties derived from Monte Carlo simulations for each of the three epochs.}
\label{fig:ijhk}
\end{figure*}

The near-simultaneous $J$, $H$, and $K_s$ light curves taken on 26 Sep 2009, 30 Sep 2009, and 01 Oct 2009 are shown in figure \ref{fig:ijhk}.  Unbinned data are shown in the top panel, while in the middle panel the data have been binned to one data point per filter change (6 exposures per epoch in $J$, and 10 in $H$ and $K_s$, less exposures discarded due to bad pixels).  Error estimates for the unbinned data points are given by the standard deviation of all measurements in a continuous segment at a given filter position, while the binned data points were assigned uncertainties of $1/\sqrt{N}$ times lower, where $N$ is the number of binned exposures at a given filter position.  Relative amplitudes of variability in the different bands were measured by assuming all light curves to have the same functional form, but with differing amplitudes.  For this purpose we used the $J$-band light curves as templates, linearly interpolating between binned data points.  Scaled versions of the $J$-band templates were then fit to the $H$ and $K_s$ light curves using a simple weighted linear regression.  The best-fit solutions (which we will refer to as the $H$ and $K_s$ band templates) are overplotted on the data in figure \ref{fig:ijhk}.  We find amplitude ratios of $A_H/A_J$=$\{$0.91$\pm0.07$, 0.84$\pm$0.08, 0.91$\pm$0.15\} and $A_{K_s}/A_J$=$\{0.83\pm0.08, 0.59\pm0.07, 0.45\pm0.11\}$ for the earliest to latest epoch respectively.  Here, $A_i$ represents a peak-to-peak amplitude of photometric variability in a bandpass $i$ given by the absolute change in brightness divided by the mid-brightness.  Uncertainties for the amplitude ratios were determined by repeating the above procedure on a set of 1000 simulated light curves.  The simulated light curves were obtained from sampling the $J$, $H$, and $K_s$ band templates at the times of our binned data points in figure \ref{fig:ijhk}.  A random noise component was added to each point, drawn from a gaussian distribution with a standard deviation equal to the 1$\sigma$ uncertainty corresponding to that data point.  In practice we found that the fits between the $H$ and $K_s$ light curves and scaled $J$-band templates produced reduced $\chi^2$ values from 1.4-3, and we therefore scaled the width of the random error component to achieve \postsub{a reduced} $\chi^2 \sim 1$ in comparison to the best-fit templates.  Measured light curve amplitudes and ratios for the different epochs are tabulated in table \ref{tab:red}. 

The results from the near-simultaneous sequences are somewhat surprising.  The light curves from the consecutive nights of Sep 21-24, where observations were made in a single but different band each night (figure \ref{fig:jhk}), would have led us to conclude that variations in the $H$ and $K_s$ bands are only 50-60\% of those in $J$.  However, amplitude ratios measured from the first near-simultaneous sequence on 26 Sep 2009 appear close to unity.  Upon repeating the near-simultaneous measurements at two subsequent epochs we continued to find $A_H/A_J \gtrsim 0.84$, while we found $A_{K_s}/A_J$ as low as $0.45$ in the latter epoch.  \colb{Our data may hint at amplitude ratios that are themselves variable, as amplitude ratios from the first and last epoch with simultaneous $JHK_s$ data are incompatible at the 2$\sigma$ level.}  In addition, none of the peak to peak amplitudes measured for the simultaneous $JHK_s$ light curves (irrespective of bandpass) are as large as the lowest amplitudes measured from the single-band consecutive night sequences of 21-24 Sep 2009.  Therefore, there is evidence that both amplitude ratios, as well as the overall amplitude may vary with either epoch, light curve phase, or a combination of both.

\subsection{Rotation and Longer Timescale Trends}
\begin{figure}[float]
\epsscale{1.2}
\plotone{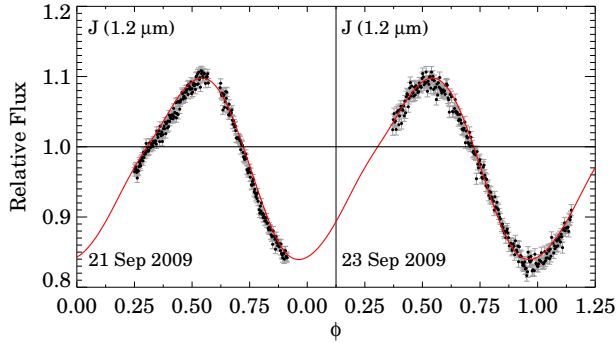}
\caption{{\em Top:} Best-fitting round,  two-spot model (red curve, P=7.721~hr) to the $J$-band light curves from 21 and 23 Sep 2009 (data points). \label{fig:2spot}}
\end{figure}

While not strictly uniform from epoch to epoch, the variations exhibited by \targ show a clear periodicity.  Since the $J$-band light curves do not seem to have evolved significantly in shape from 21 Sep 2009 to 23 Sep 2009 we determined an approximate period by fitting a periodic function to these light curves.  Since variations are not perfectly sinusoidal, we chose a  round,  non-overlapping, two-spot model \citep{dorren87} as our periodic function, and performed fitting using a Markov Chain Monte Carlo \citep[e.g.][]{lewis02} technique with a Metropolis Hastings algorithm \citep{metropolis53,hastings70} (figure \ref{fig:2spot}).  Using this method we determined a period of  \postsub{$7.721\pm0.005$ hr}, with the quoted value and errors derived from the maximum likelihood and 67\% credible regions of the posterior distribution of periods.  \postsub{The uncertainty is likely underestimated due to the implicit assumption that the light curve shape has not evolved between cycles used for fitting.}
 
Two spots are required to fit the asymmetry in the light curve, which is hinted at from a secondary ``bump'' present at the beginning of the $J$-band light curve from 21 Sep 2009, and which also becomes apparent when observations are phased to the best-fit period.  

\begin{figure}[float]
\epsscale{1.2}
\plotone{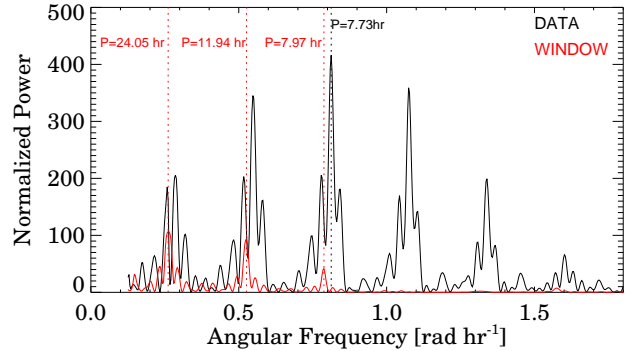}
\caption{Lomb-Scargle Periodigram (black line) of  light curves spanning 21 Sep 2009 to 01 Oct 2009.  
The power corresponding to a false alarm probability of 0.997 is $\sim$4, falling well below any peaks in the periodigram.  
The scaled window function (red line) shows a main peak at 24~hr corresponding to the spacing of our nightly observations, with additional peaks at 12~hr, 8~hr, etc.  The strongest data peak at 7.73~hr matches the 7.721$\pm0.005$~hr value recovered from fitting a periodic function to the light curve (see figure \ref{fig:2spot}). \label{fig:lsper}}.
\end{figure}

In addition to fitting for the period, we also computed a Lomb-Scargle Periodogram \citep{scargle82,horne86} of the $J$-band data spanning  21 Sep 2009 to 01 Oct 2009 (figure \ref{fig:lsper}).  In order not to favor epochs with more densely sampled data, all light curves were linearly resampled at 10-min spacing within each observing window. Despite falling close to the 8~hr sub-peak of the window function (which peaks at 24~hr), the strongest peak at 7.73~hr matches the \postsub{7.721$\pm0.005$~hr} recovered from our spot modeling.

\begin{figure}[float]
\epsscale{1.1}
\hspace{0cm}
\plotone{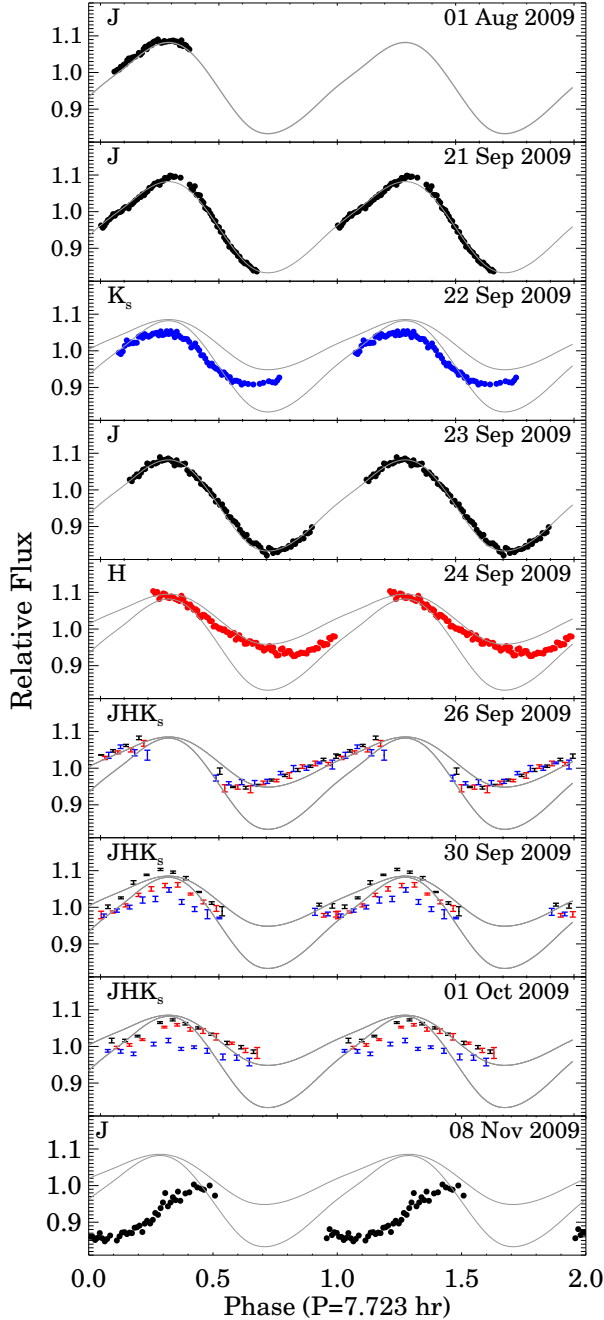}
\caption{$J$ (black) $H$ (red), and $K_s$ (blue) light curves spanning 01 Aug 2009 to 08 Nov 2009 shown on a common flux scale where unity corresponds to $J_{MKO}$, $H$, and $K_s$ magnitudes of 14.75, 14.11, and 13.59 respectively.  \colj{The data have been phased to an over-constrained period of 7.723~hr  (see text for the explanation).  For clarity, the data have been binned by factors of 3 (WIRC $J$-band sequences),  7 (CPAPIR $J$-band sequence in bottom panel), and 5 ($H$ and $K_s$ sequences).  The interleaved $JHK_s$ sequences are binned as in figure \ref{fig:ijhk}.
The best-fitting two-spot model from figure \ref{fig:2spot} is overplotted (solid grey lines) to \colj{highlight deviations from the original light curve shape as a function of time}.  The same model, with the amplitude scaled by 55\%, is also overplotted on the lower amplitude sequences as a visual aid. \label{fig:jphase}}}
\end{figure}

We find that the uncertainty in the best-fit rotation period of \postsub{$7.721\pm0.005$~hr} gives us the \postsub{freedom to phase the Aug 2009 light curves with those from Sep 2009, but there is no single period that can accommodate all epochs spanning Aug 2009 - Nov 2009.}  \colj{As an example of this}, all light curves (presented on a common flux scale) are shown in figure \ref{fig:jphase}, phased to an over-constrained period of 7.723~hr.   In fine-tuning the period to precisely 7.723~hr, we have arbitrarily chosen to phase the earlier (Aug 2009) observations with those from Sep 2009 \colj{in order to illustrate the resultant mismatch in phases for the later (Nov 2009) epoch.  However, given the lack of a common phasing across all epochs, there is no reason to expect the 01 Aug 2009 observations to be in phase with the middle epochs. Thus, while we have used a period of precisely $P=7.723$~hr in our figures, it should be understood that this is illustrative, and that the actual periodicity cannot be constrained beyond \postsub{$7.721\pm0.005$~hr} from our observations.}

While the maxima and minima of the light curves spanning 21 September 2009 to October 01 2009 appear roughly synchronized with a 7.723~hr period, there are large differences in amplitude, and more subtle differences in light curve shape from epoch to epoch.  

\begin{figure}[float]
$\begin{array}{c}
\includegraphics[width=3.2in]{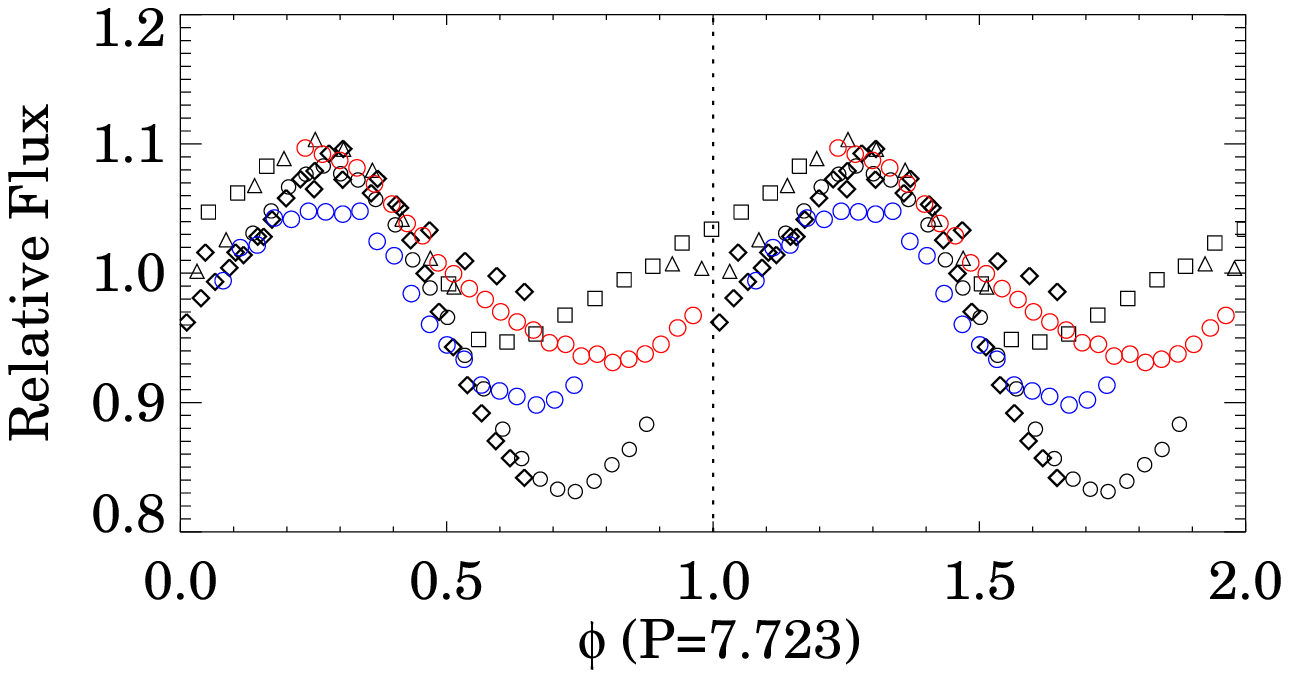} 
\\
\includegraphics[width=3.2in]{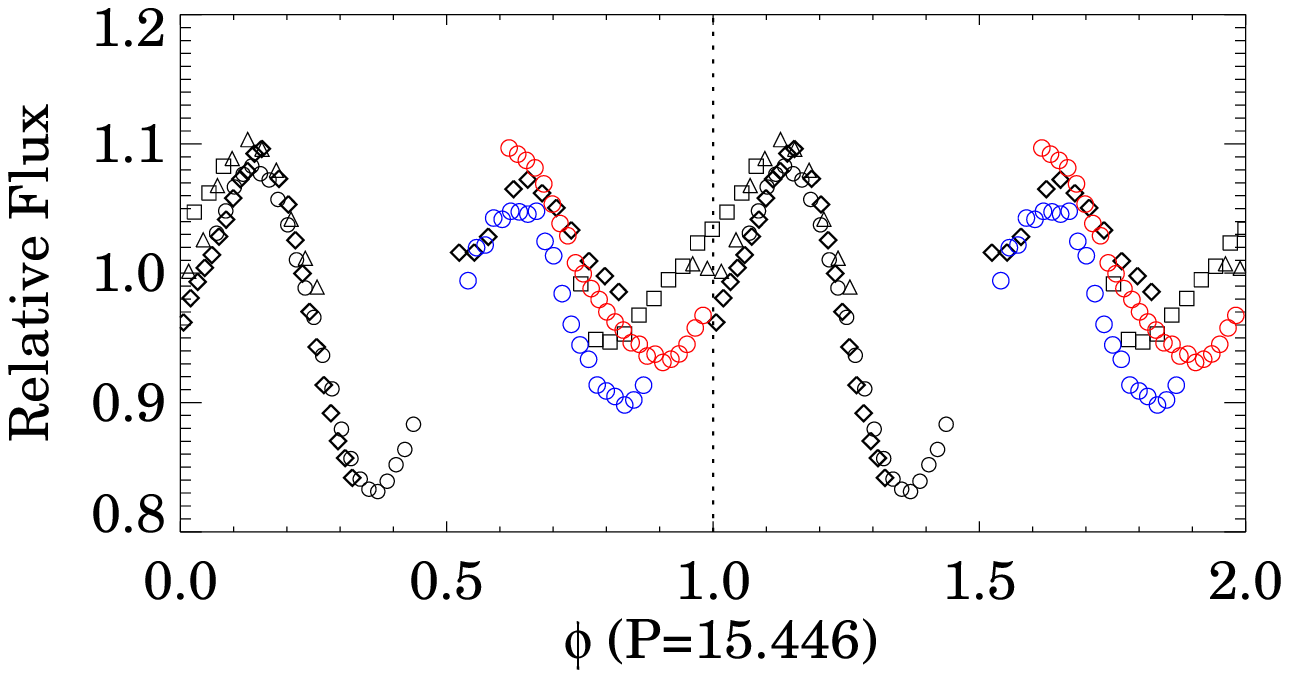}
\end{array}$
\caption{Light Curves for \targ spanning 21 Sep 2009 to 01 Oct 2009 phased to over-constrained (see text for explanation) periods of 7.723~hr (top) and 15.446~hr  (bottom). Black, red and blue points represent $J$, $H$, and $K_s$ band data respectively.  Different symbols of the same color differentiate different epochs.  The data have been binned for clarity.  Unity corresponds to $J_{MKO}$, $H$, and $K_s$ magnitudes of 14.75, 14.11, and 13.59 respectively. \label{fig:dblpk}}
\end{figure}

There is a marked decrease in the observed light curve amplitudes between the 23 Sep 2009 and 26 Sep 2009 epochs.  While this may indicate short timescale evolution of cloud features, it is also consistent with a more stable, double-peaked light curve with a period of 2$\times$7.72=15.44~hr such that the higher-amplitude variations observed on 21 Sep 2009 and 23 Sep 2009 sample a different phase of the light curve than the lower-amplitude variations observed at subsequent epochs.  This possibility is illustrated in figure \ref{fig:dblpk} where light curves spanning 21 Sep 2009 to 01 Oct 2009 are shown phased to both 7.723~hr and 15.446~hr (once again, over-constrained) periods.  For the case where $P=15.446$~hr, overall changes in light curve shape and amplitude, although still apparent, are less pronounced.  A period of 15.44~hr would make \targ a \colb{somewhat slow rotator} in comparison to other ultracool dwarfs \colb{with $v\sin{i}$ measurements}, which have inferred periods ranging from $\sim$2-12~hr \citep[e.g.][]{reiners08}. 

\colj{There appears to be some phase coherence on a timescale of days to weeks, suggesting that we are observing cloud features that persist at least this long.
Over months the phase coherence is lost, which may indicate evolution of the cloud coverage or the dissipation and formation of new features.   An intriguing possibility is that the long-timescale evolution of the light curve is caused by the differential rotation of a storm system with respect to cloud features at different latitudes \citep[e.g.][]{artigau09}.  The light minimum of the last epoch (8 Nov 2009) is approximately 3~hr ahead of the continuous $J$- band sequence from 23 Sep 2009, which could imply a differential rotation of $\sim$3 degrees per day or wind speeds of $\sim$45~m~s$^{-1}$ during that time period (approximately half of that on Jupiter).  We note that this estimate relies on our over-constrained period of 7.723 hr which was obtained by assuming a {\em constant} phase for the Aug through Oct epochs, which may not be realistic.  
Long term monitoring of \targ should be able to verify the existence of a persistent, differentially rotating feature.}

\begin{deluxetable*}{lcccccc}
\tablecaption{Light Curve Properties \label{tab:red}}
\tablecolumns{7}
\tablewidth{7in}
\tablehead{ 
\colhead{Epoch\tablenotemark{a}} &  \colhead{Filter}  & \colhead{$A$\tablenotemark{b}} & \colhead{$A_H/A_J$} & \colhead{$A_{K_s}/A_J$} & \colhead{\colj{$\sigma_t$}} & \colhead{\colj{$\sigma_{c}$}}
}
\startdata
01 Aug 2008 & $J$ & 0.080 & -- & -- & 0.0072 & 0.0056\\
21 Sep 2009 & $J$ &  0.260 & -- & -- & 0.0067 & 0.0051 \\ 
22 Sep 2009 & $K_s$  &  0.156 & -- & -- & 0.0145 & 0.0156 \\
23 Sep 2009 & $J$ &  0.262 & -- & -- &  0.0079 & 0.0070\\
24 Sep 2009 & H & 0.166 & -- & -- & 0.0124 & 0.0119\\
26 Sep 2009 & $J$ ($H$,$K_s$) & 0.134 (0.123, 0.111) & 0.91$\pm$0.07 & 0.83$\pm$0.08 & 0.0092 (0.0159,0.0186) & 0.0088 (0.0136,0.0236) \\
30 Sep 2009 & $J$ ($H$,$K_s$) & 0.109 (0.092, 0.064) & 0.84$\pm$0.08 & 0.59$\pm$0.07 & 0.0107 (0.0154,0.0201) & 0.0107 (0.0132,0.0227) \\
01 Oct 2009 & $J$ ($H$,$K_s$) &  0.084 (0.077, 0.038) & 0.91$\pm$0.15 & 0.45$\pm$0.11 & 0.0088 (0.0142,0.0175) & 0.0076 (0.0116,0.0188) \\
08 Nov 2009 & $J$ &  0.152 & -- & -- & 0.0391 & 0.0232 
\enddata
\tablenotetext{a}{As in table \ref{tab:log}, dates correspond to the local day at the beginning of the night.}
\tablenotetext{b}{Peak-to-peak amplitudes are measured as the absolute change in flux, divided by the mid-flux and correspond to the filter(s) indicated in the second column.}
\end{deluxetable*} 

\subsection{Possible long-term variability}
From comparing our WIRC photometry to that from 2MASS (both on the 2MASS photometric system, see table \ref{tab:phot}), we find \targ to be significantly bluer ($J-K_s$=1.26-1.34) at the time of the WIRC observations than reported by 2MASS ($J-K_s$=1.67$\pm$0.07).  We can estimate \targs's maximum change in $J-K_s$ color within the time spanned by our WIRC observations by making the conservative assumption of $A_{K_s}/A_J$=0.45 (see figure \ref{fig:ijhk} and table \ref{tab:red}).  This amplitude ratio, the lowest in our data, can be used to infer a maximum $\Delta (J-K_s)$ of 0.15 mag for the full range of $J$ band variability observed ($A_J$=0.26).  Therefore, there is no way to reconcile the 2MASS catalog color of $J-K_s$=1.67$\pm$0.07 with that observed by WIRC, unless large systematic errors exist in our photometric calibration.  Fortunately, NIR spectra for \targ from the SpeX Prism Library provides two additional epochs which we can compare to WIRC and 2MASS photometry.  Synthetic 2MASS colors of $J-K_s$=1.32$\pm0.083$ and 1.36$\pm0.080$ were found for the 2003 and 2004 SpeX epochs respectively, in agreement with the WIRC photometry.  It is interesting to note that compared to 198 other L and T dwarfs for which we have also measured synthetic SpeX photometry (see Appendix \ref{sect:ap1}), \targs's large offset between 2MASS and SpeX colors makes it a $4\sigma$ outlier.  Thus it may have been possible to identify this target's variability well in advance of our observations.  It is also notable that there is no apparent increase in scatter between 2MASS and SpeX photometry across the L/T transition, with a few other less extreme outliers occurring at earlier spectral types.  These additional outliers should be followed up, although it appears that variability on the scale observed for \targ is quite rare.

Thus, if the 2MASS epoch can be trusted, there is evidence that the entire range of \targs's variability is larger than captured by our observations.  We note that there is no reason to suspect an error in the 2MASS colors, as catalog magnitudes derived from profile fitting agree well with those from aperture photometry, and the 2MASS $J$, $H$, and $K_s$ observations are simultaneous.  Furthermore,  according to our WIRC photometry the noted blueward shift in $J-K_s$ since the 2MASS epoch is almost entirely attributable to a brightening in the $J$-band, accompanied by a plateau or possible dimming in the $K_s$ band.  If real, this long-term trend of anti-correlated behavior in $J$ and $K_s$ is qualitatively different than the highly correlated variability we observed over short timescales.

\section{Physical Properties}
\label{sect:phys}
The physical properties of \targ derived here and elsewhere are presented in table \ref{tab:phot}.  \targ is a cool field BD with an optical spectral type of T0 determined by \citet{reid08}, and a NIR spectral type of T1.5 from \citet{burgasser06}.  The NIR spectra (R$\sim$120) from \citet{burgasser06} were taken with the Medium Resolution Near Infrared Spectrograph (SpeX) at the NASA's InfraRed Telescope Falcility (IRTF), and are available online from the SpeX Prism Library.  There is no parallax data available for this target at this time, and given it's unique variability (and hence potentially unique atmospheric characteristics) it is unclear to what degree standard relationships between spectral type, absolute magnitude and temperature may apply.  \colj{For field BDs ($\sim$3~Gyr) with measured parallaxes and estimated bolometric luminosities it has been shown that effective temperatures plateau or slightly decrease from $\sim$1400-1200~K across the L/T transition \citep[$\sim$L7-T4 SpTs;][]{golimowski04,stephens09}. The scatter about this temperature is reported as $\sim$100~K.  Using the relationship provided by \citet{stephens09} we derive $T_{eff}$=1270$\pm$100\s K for a T1.5 spectral type.    However, the temperature of the L/T transition also shows some dependence on gravity \citep[e.g.][]{metchev06,leggett08,stephens09}, and this temperature could be lower if \targ is moderately young.}

If we assume 2M2139 to have an age/mass typical of field T dwarfs, an absolute magnitude and distance can be determined from empirical relationships.  \colj{Using the spectral type versus MKO $K$ magnitude relation given by \citet{marocco10}, excluding known binaries and using a NIR spectral type of T1.5, we find an absolute magnitude of $M_K=13.35\pm0.25$~mag.  The error bar reflects the coefficient uncertainties of the polynomial fit provided by \citet{marocco10}, \postsub{but neglects intrinsic scatter of the sample about the relation.  This intrinsic scatter is estimated by \citet{liu06} to be $\sim$0.39\s mag, and we therefore adopt this larger uncertainty hereafter.}  Assuming \targ is a single object, this corresponds to a distance of \postsub{$11.6^{+2.3}_{-1.9}$\s pc}, where we have used the 2MASS $K_s$ magnitude of 13.58~mag, first converted to an MKO K magnitude using the relationship provided by \citet{stephens04}, to obtain the distance modulus.}   Our distance estimate is notably closer than that of 18.8~pc cataloged by \citet{looper07} using the MKO-J relationship of \citet{liu06}.

\begin{deluxetable}{lccc}
\tablecaption{Target Properties\label{tab:phot}}
\tablecolumns{4}
\tablewidth{3.4in}
\tablehead{
\colhead{Quantity} & 
\colhead{Value or Range} & 
\colhead{Data Ref.} & 
\colhead{Source}}
\startdata
Identifier &  J21392676+0220226 & 2MASS\tablenotemark{a}& 2\\
$\alpha$ (J2000)  & 21$^h$39$^m$26$^s$.76 & 2MASS\tablenotemark{a}& 2\\
$\delta$ (J2000)  & +02$^{d}$ 20\arcmin22\arcsec .6  & 2MASS\tablenotemark{a}& 2\\
Optical SpT & T0 & -- & 3\\
NIR SpT & T1.5 & -- & 4\\
Period & \postsub{7.721$\pm$0.005~hr} & --& 1\\
T$_{eff} $ (3 Gyr) & $1270\pm100$~K \tablenotemark{e}& --& 1\\
$M_{K,MKO}$ (3 Gyr) & $13.35\pm0.25$ \tablenotemark{f}& --  & 1\\
\colj{$d$ }& \postsub{$11.6^{+2.3}_{-1.9}$~pc\tablenotemark{f}} & --& 1\\
\tableline
\multicolumn{4}{c}{Photometry}\\
\tableline
$J$ & 15.26$\pm$0.049 & &  \\
$H$ &   14.16$\pm$0.053  & 2MASS\tablenotemark{a}& 2\\
$K_s$ &  13.58$\pm$0.045  & & \\
$J-K_s$ &  1.67$\pm$0.066  & & \\
$J-H$ &  1.10$\pm$0.072  & & \\
\tableline
$J-{K_s}$ & 1.32$\pm$0.07  & \\
$J-{H}$ &  0.89$\pm$0.07  & \\
$\Delta J_{(2M-MKO)}$ & 0.183  &SpeX\tablenotemark{b}& 1\\
$\Delta H_{(2M-WIRC)}$ & -0.065 & & \\
$\Delta {K_s}_{(2M-WIRC)}$ & -0.016  & & \\
\tableline
$J_{MKO}$ &  14.66-14.96\tablenotemark{c} &  & \\
$J$ &  14.84-15.14\tablenotemark{c}  & & \\
$H$ & 14.00-14.18\tablenotemark{c} & WIRC & 1\\
$K_s$&  13.54-13.71\tablenotemark{c} & & \\
$J-K_s$&   1.26-1.34\tablenotemark{d}  & & \\
$J-H$ &  0.78-0.83\tablenotemark{d} & & 
\enddata
\tablenotetext{a}{2MASS Point Source Catalog, the epoch is JD 2451741.8470 (16 Jul 2000)}
\tablenotetext{b}{Synthetic photometry using 2M2139's  NIR spectrum \citep{burgasser06} from the SpeX Prism Library.  Note that the closely agreeing 2003 and 2004 epochs have been averaged.}
\tablenotetext{c}{Ranges span the entire set of WIRC observations (i.e. all epochs) presented here.}
\tablenotetext{d}{Ranges correspond only to epochs where near-simultaneous $JHK_s$ photometry is available.}
\tablenotetext{e}{Derived from the empirical relation of \citet{stephens09}}
\tablenotetext{f}{Derived from the empirical relation of \citet{marocco10}}
\tablecomments{References refer to (1)This paper, (2)\citet{2mass}, (3)\citet{reid08}, (4)\citet{burgasser06}}
\end{deluxetable}

\subsection{Binarity?}
\citet{burgasser10} have suggested that \targ is a binary candidate since a composite spectral template consisting of L8.5$\pm0.7$ and T4.5$\pm1.5$ components provide a significantly better fit to its NIR spectrum than any single template.  For example, pronounced CH$_4$ absorption is present at 1.2 $\mu$m, but is weak or absent in $H$ and $K_s$.  One way to reproduce this mismatch is with the addition of a late T companion with highly suppressed $H$ and $K_s$ fluxes;  this would enhance the height of the $J$-band peak,  while having a much smaller effect on the $H$ and $K$ band SEDs.  However, even this composite match fails to reproduce the relative strength of CH$_4$ absorption in the $J$, $H$ and $K_s$ bands, leading the authors to conclude that if a binary, the components ``may themselves have unusual properties.''  Given the unique variability of \targ reported in the present work, its peculiar NIR spectrum may be the result of an uncommonly heterogeneous atmosphere, rather than binarity. 

2MJ2139 was observed on 2006 June 23 using HST/NICMOS with the NIC1 camera (43 mas pixel$^{-1}$) and the F170M and F110W filters (HST program ID 10143).  We analyzed these data to put constraints on the binarity of this brown dwarf.  For point-spread function (PSF) fitting and subtraction, we used both a model PSF generated with the TinyTim\footnote{\url{http://www.stsci.edu/hst/observatory/focus/TinyTim}} software \citep{krist93} as well as the image of another brown dwarf (2MJ0257-3105) observed with the same settings only two days later as part of the same program. The latter reference PSF provided a noticeably better fit to the image of 2MJ2139+02 than the former, although in both cases the subtraction left no significant residuals. Thus the data provide no indication of binarity. By introducing fake binary companions of various contrasts and separations in the images and repeating the reference PSF fitting, we can rule out the presence of a binary companion with contrasts of 0, 1, 2, and 3 mag at separations larger than 0.055\arcsec, 0.065\arcsec, 0.08\arcsec and 0.13\arcsec, respectively. These limits correspond to physical separations between 0.66-1.56~AU at a distance of 12~pc.

Given the apparent proximity of this source, future parallax measurements should provide more stringent constraints on its distance, absolute magnitude, and potential binarity.

\subsection{Model Fits to the NIR SED}
\label{sect:single}
Detailed atmosphere models including the effects of dust condensation and settling can be used to explore atmospheric properties of field BDs including effective temperature, gravity, metallicity, vertical mixing (via departures from chemical equilibrium), and condensate clouds \citep[e.g.][]{cushing08,stephens09,witte11}.  However, 
certain issues with atmosphere models are well documented.  For instance, studies of benchmark objects---those with known masses \citep[e.g.][]{dupuy11}, and/or ages \citep{leggett08}---have demonstrated that effective temperatures determined from model atmosphere fitting can differ by up to a few hundred degrees compared to those derived from evolutionary models.  \colj{Second, low-gravity L dwarfs (Faherty et al. 2011, submitted) and directly imaged planets \citep[e.g.][]{bowler10,skemer11} have been found to be less luminous than ``normal'' field BDs of a given spectral type.  This latter observation is surprising as lower gravity objects are expected to be larger in radius, and hence brighter.  The ability of low gravity atmospheres to retain a thicker and higher cloud layer, high metallicity, and/or non-equilibrium chemistry may be responsible \citep[e.g.][]{barman11}, and highlights that significant interdependencies between effective temperature, condensate properties, surface gravity, and other secondary parameters may exist.}

In addition to known problems, it is unclear whether standard 1D models are appropriate for \targ as it's large variability indicates a heterogeneous surface.  Nonetheless, we have performed fits of 1D model atmospheres to \targ's NIR spectrum in order to determine the general properties required to reproduce its spectrum in a manner consistent with other studies.  Also, in the following section we model \targ's variability and NIR spectrum simultaneously using linear combinations of 1D cloudy and clear model atmospheres.  Thus, the single spectral fits performed here will serve as a reference to which hybrid cloudy/clear models can be compared.

\begin{figure*}
\epsscale{0.95}
\plotone{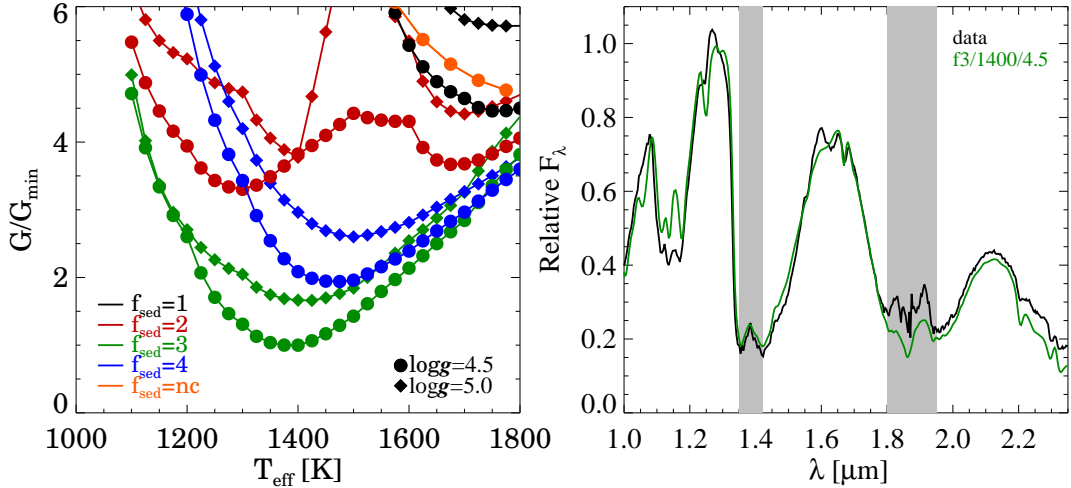}
\caption{{\em Left panel:} The weighted $\chi^2$ statistic $G$ (normalized to its overall minimum value, $G_{min}$) for a series of different $f_{\rm sed}$, $\log{g}$, and $T_{eff}$. {\em Right panel:} The model spectrum corresponding to the minimum $G$ statistic (green line) is shown overplotted on the data (black line).  Regions of high telluric absorption which have been omitted from the fit are shown in grey. \label{fig:single}}
\end{figure*} 

\subsubsection{The Atmosphere Models}
The model spectra employed here have been calculated for solar metallicity and chemical equilibrium, for simplicity allowing only effective temperature ($T_{eff}$), surface gravity ($\log{g}$) and cloud sedimentation efficiency ($f_{\rm sed}$) to vary.   The $f_{\rm sed}$ parameter describes the efficiency of condensate sedimentation within the atmosphere, according to the cloud model of \citet{ackerman01}, which has been successfully used to reproduce observations of a wide range of cloudy BDs \citep[e.g.][]{cushing08,stephens09} as well as Jupiter.  Large values of $f_{\rm sed}$ correspond to more efficient particle growth, resulting in larger particle sizes that more efficiently rain out of the atmosphere and consequently lower opacity cloud decks.  Our model grid was calculated for 500~K$<T_{eff}<$2000~K in steps of 100~K, $\log{g}=$\{4.5,5.0\}, and $f_{\rm sed}$=\{1,2,3,4,nc\}, where `nc' denotes a cloudless model.  Although the models were calculated with 100~K spacing, we have linearly interpolated between them to obtain a finer sampling in $T_{eff}$ of 25~K.  

\subsubsection{Fitting procedure}
For each model in our grid we computed a weighted $\chi^2$ statistic, following a modified version of the procedure described by \citet{cushing08}, and given by

\begin{equation}
G = \sum_{i}w(\lambda_i) \left(\frac{F(\lambda_i) - c M(\lambda_i;T_{eff},\log{g},f_{\rm sed})}{\sigma_i}\right)^2
\label{eq:fit}
\end{equation}

\noindent where $F(\lambda_i)$ is the observed spectrum, and $M(\lambda_i;T_{eff},\log{g},f_{\rm sed})$ is the model, scaled by a constant factor $c\equiv(R/d)^2$ that minimizes $G$, where $R$ is the BD radius and $d$ is its distance. All model spectra were smoothed to replicate the average resolution of \targs's spectrum over the wavelength region of interest (R$\sim$100), via convolution with a Gaussian.  Due to unequal spacing of the data, the fit is weighted by $w(\lambda_i)=\Delta \lambda_i$, where $\Delta \lambda_i$ is the width of wavelength bin $i$.  Fitting was performed over the wavelength intervals of 1.0-1.35 $\mu$m, 1.42-1.8 $\mu$m, and 1.95-2.35 $\mu$m, avoiding regions of high telluric water absorption.  Due to uncertainties in molecular band strengths and opacities, as well as the complexity of condensate cloud physics, the models are generally incomplete and can disagree with the observations at a level significantly greater than the measurement uncertainties associated with the data.  Thus we have chosen to set all $\sigma_i=1$, so as not to bias a particular wavelength regime based on their measurement uncertainties.  This is equivalent to including a constant uncertainty term for each wavelength bin in order to account for model incompleteness.  Since this constant term is much larger than the measurement uncertainty for any given wavelength bin included in the fit, the $\sigma_i$ become essentially constant.  

In figure $\ref{fig:single}$ we plot the $G$ statistic (normalized by its absolute minimum value across all models tested, $G_{min}$) for various values of $f_{\rm sed}$ and $\log{g}=\{4.5,5.0\}$ as a function of effective temperature.  We find the best fit corresponds to $\log{g}=4.5$, $f_{\rm sed}$=3 and $T_{eff}$=1400~K.  While an effective temperature of 1400~K is consistent with the range of L/T transition temperatures found by \citet{golimowski04} and \citet{stephens09}, a surface gravity of $\log{g}$=4.5 would then imply a rather young age of 100~Myr and mass of $\sim$20 $M_{Jup}$ \citep{saumon08}. 
\colj{It is interesting to note that the $K$ band portion of \targs's SED has been reported as a best match to that of the low gravity directly imaged planet  HR8799~b \citep{barman11}.  However, model fits to the low resolution NIR spectra of field BDs do not strongly constrain surface gravity, and can yield values of $\log{g}$ differing by up to 1.0 dex when different wavelength regimes are used for fitting \citep{cushing08}.  In addition, according to \citet{cushing08} a value of $\log{g}=4.5$ derived from spectral fitting is not particularly unusual for a field BD, even when evolutionary sequences predict higher surface gravities ($\log{g}>5$) for the same objects.  Thus, it would be premature to conclude anything about \targ's age and gravity based on model fits alone, especially since there is no additional evidence for youth.  With this in mind, it may be more likely that interdependencies between surface gravity and other physical parameters such as cloud thickness and/or metallicity (despite being assigned independent model parameters) lead to the latter being mimicked by the former in spectral models.  For the purpose of this paper we continue to use the $\log{g}=4.5$ models to describe \targ's atmosphere, but caution that the physical interpretation of the gravity parameter is unclear.  The analysis presented here and in subsequent sections is not strongly affected by this choice of model surface gravity, other than providing an optimal fit to \targ's NIR SED.  Specifically, trends pertaining to photometric variability in the following sections are qualitatively similar for different surface gravities. Furthermore, from figure $\ref{fig:single}$, $T_{eff}$ appears roughly independent of $\log{g}$, but depends strongly on $f_{\rm sed}$.  Thus, while $\log{g}$ is poorly constrained by the models, it is only of secondary importance to our analysis in comparison to the cloud thickness parameter, $f_{\rm sed}$.  }

\section{Modeling Variability due to Heterogeneous Surface Features}
\label{sect:model}
While we discuss alternative explanations for \targs's variability in the discussion section, here we construct a simple model for the observed variability, assuming that heterogeneous surface features are responsible.  We envision a scenario where ``spots'' or cloud features remain approximately static in the BD's rotating frame over a single rotation, and variability arises due to rotational modulation.  We consider a simple model where the BD surface is composed of two types of regions differing in temperature and/or cloud properties, possessing surface fluxes $\mathcal{F}_1$ and $\mathcal{F}_2$.  At a given snapshot in time the total flux from the BD, $\mathcal{F}$, is given by a linear combination of $\mathcal{F}_1$ and $\mathcal{F}_2$ weighted by their relative filling fractions over the BD's visible disc.  The peak-to-peak amplitude of variability that an observer would detect in a given bandpass due to a change in these filling factors can be expressed as

\footnotesize
\begin{equation}
\label{eq:var1}
A=\frac{(1-a-\Delta a)\mathcal{F}_1 + (a+\Delta a)\mathcal{F}_2 - (1-a)\mathcal{F}_1 - a\mathcal{F}_2}{\colb{0.5[(1-a-\Delta a)\mathcal{F}_1 + (a+\Delta a)\mathcal{F}_2 + (1-a)\mathcal{F}_1 + a\mathcal{F}_2 ]}}
\end{equation}
\begin{equation}
  \label{eq:var2}
 = \frac{\Delta a}{\colb{\alpha}+\mathcal{F}_1/\Delta \mathcal{F}}
\end{equation}
\normalsize

where $A=\Delta{\mathcal{F}}/\mathcal{F}$ is the change in flux divided by the mid-brightness flux.   The parameter $a$ is the minimum filling factor of the $\mathcal{F}_2$ regions, $\Delta a$ is the change in filling factor,  and $\Delta \mathcal{F}=\mathcal{F}_2-\mathcal{F}_1$.  \colb{The parameter $\alpha$=$a$+0.5$\Delta a$ corresponds to the filling factor of the $\mathcal{F}_2$ regions at mid-brightness.}

Synthetic photon fluxes $\mathcal{F}_2$ and $\mathcal{F}_1$ for use in equation \ref{eq:var2} were computed from the 1D model spectra integrated over the WIRC system plus filter transmission curves.  For this purpose we used the  models of \citet{saumon08} described above with solar metallicity, $\log{g}=4.5$, and a range of values in $f_{\rm sed}$ and $T_{eff}$.  We also experimented with different $\log{g}$ values, but found them to be a poor match to \targ's NIR spectrum, and hence only present the optimal case where $\log{g}$=4.5 here.  In general, $\mathcal{F}_1=\mathcal{F}_1[T_1, f_{\rm sed1}]$ and $\mathcal{F}_2=\mathcal{F}_2[T_1+\Delta T, f_{\rm sed2}]$ so that each region is characterized by a distinct effective temperature and cloud sedimentation efficiency.  \colb{As a matter of convention we will always choose $f_{\rm sed1}\le f_{\rm sed2}$, such that the $\mathcal{F}_1$ regions have higher condensate opacity.} The parameter $T_1$ represents the effective temperature of the $\mathcal{F}_1$ regions,  while $T_1+\Delta T$ is the effective temperature of the less cloudy $\mathcal{F}_2$ regions.  For a given combination of $T_1$, $\Delta T$, $f_{\rm sed1}$, $f_{\rm sed2}$, $a$ and $\Delta a$, fluxes can then be determined from 1D model atmosphere grids, and the resultant variability computed via equation \ref{eq:var2}.    
We caution that, in general, 1D models with different effective temperatures and vertical distributions of dust condensates have different underlying pressure-temperature profiles, and thus interpolations between them cannot be made in a self-consistent way.  Nonetheless, in the absence of 3D models, it is instructive to use 1D models as a guide, understanding their drawbacks.  Limitations of our modeling approach are discussed further in section \ref{sect:cons}.

Using equation \ref{eq:var2} we computed $A_{K_s}/A_J$ and $A_{H}/A_J$ as a function of $\Delta T$ for a variety of combinations of $f_{\rm sed1}$ and $f_{\rm sed2}$.   The various combinations can be broadly divided into \colb{two main cases}, which we specify for later reference:  

\paragraph{Case $A$} Case $A$ corresponds to an atmosphere with heterogeneous cloud features, where $f_{\rm sed1} \neq f_{\rm sed2}$.  This case could describe an atmosphere composed of clouds and clearings, or more generally regions of differing condensate opacities.  This picture of a brown dwarf with  spatially variable cloudiness is motivated by the cloud fragmentation hypothesis of the L/T transition \citep{ackerman01,burgasser02_lt}, discussed in the introduction.

\paragraph{Case $B$} Case $B$ corresponds to a uniformly cloudy atmosphere where $f_{\rm sed1} = f_{\rm sed2}$, with heterogeneities in temperature only.  This case may approximate the presence of cool or hot magnetically induced spots.  We have not modeled hot or cool spots in a uniformly {\em clear } atmosphere because the red $J-K_s$ color of \targ is incompatible with cloud-free models and precludes this scenario.  For this case we have chosen to consider only $f_{\rm sed1}=f_{\rm sed2}=3$ based on the best-fitting 1D model atmosphere for \targ (figure \ref{fig:single}).  

\begin{figure*}
\hspace{1.5cm}
$\begin{array}{c}
\includegraphics[width=2.8in]{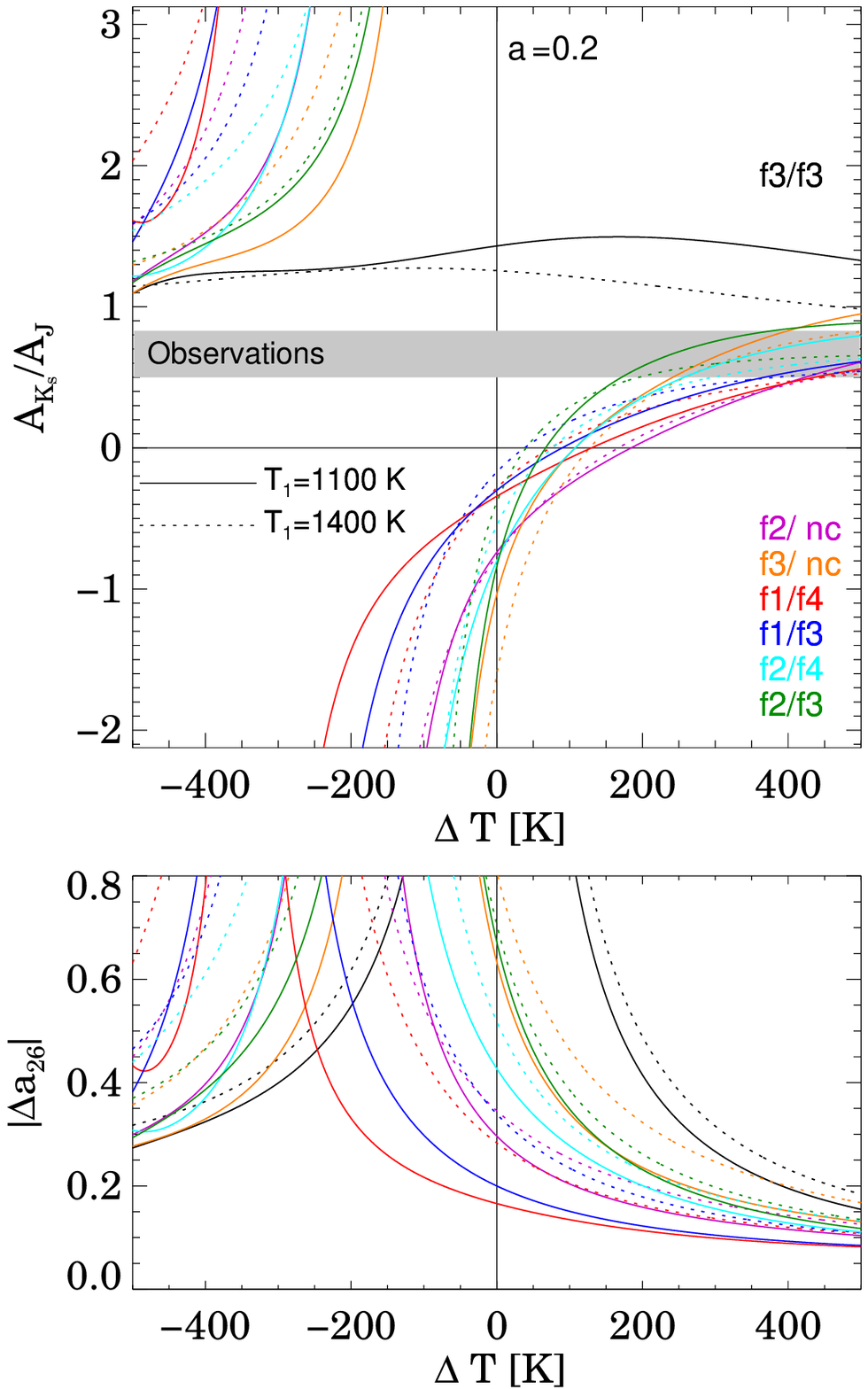} 
\includegraphics[width=2.8in]{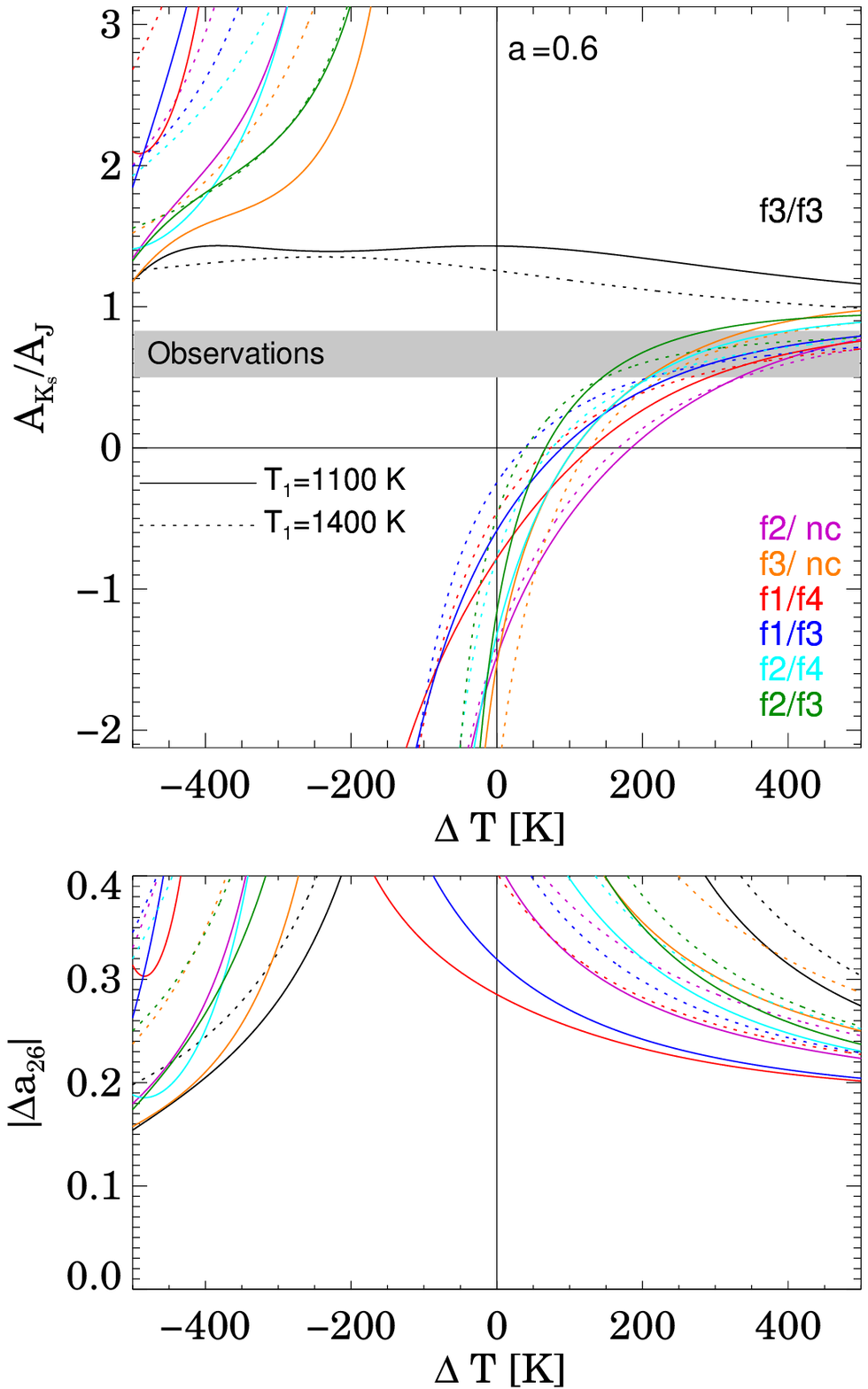} \\
\includegraphics[width=2.8in]{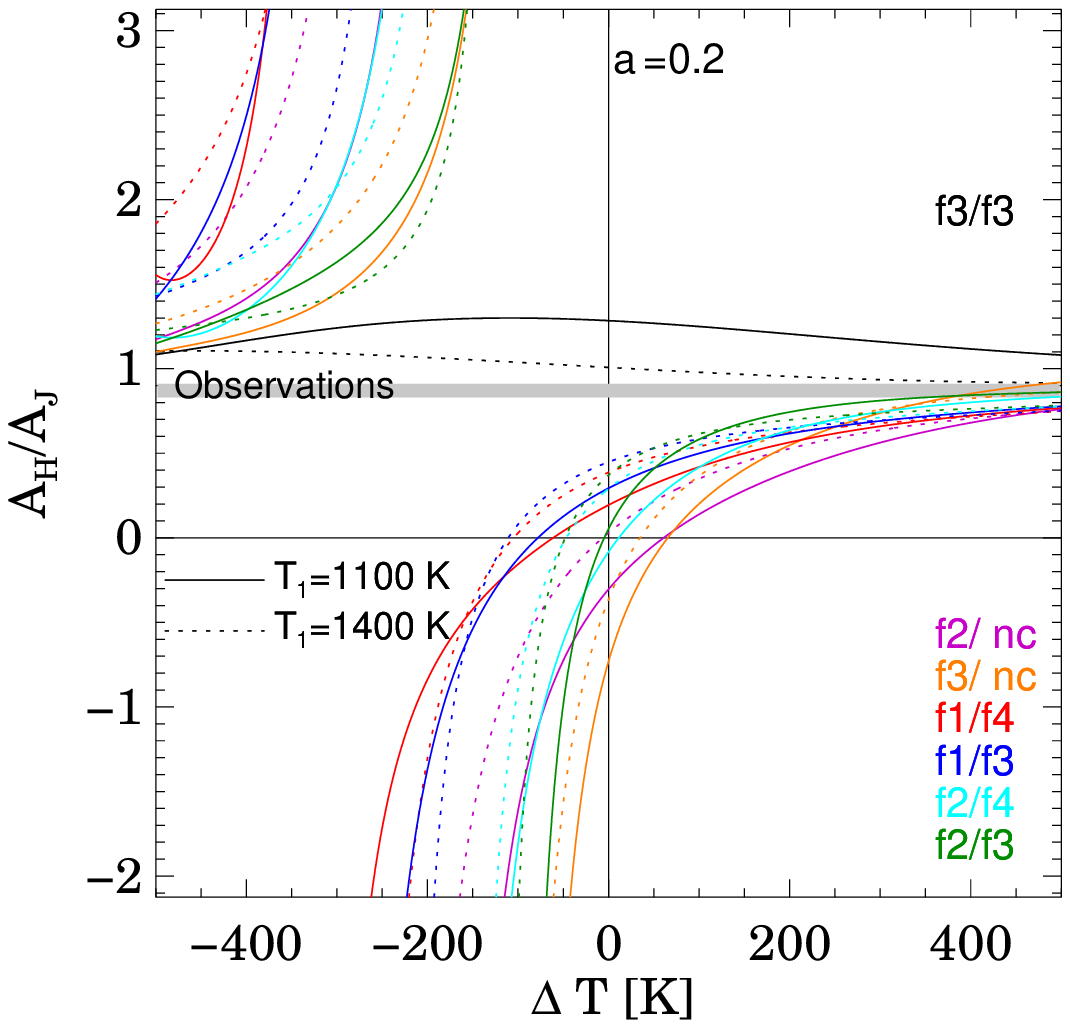}
\includegraphics[width=2.8in]{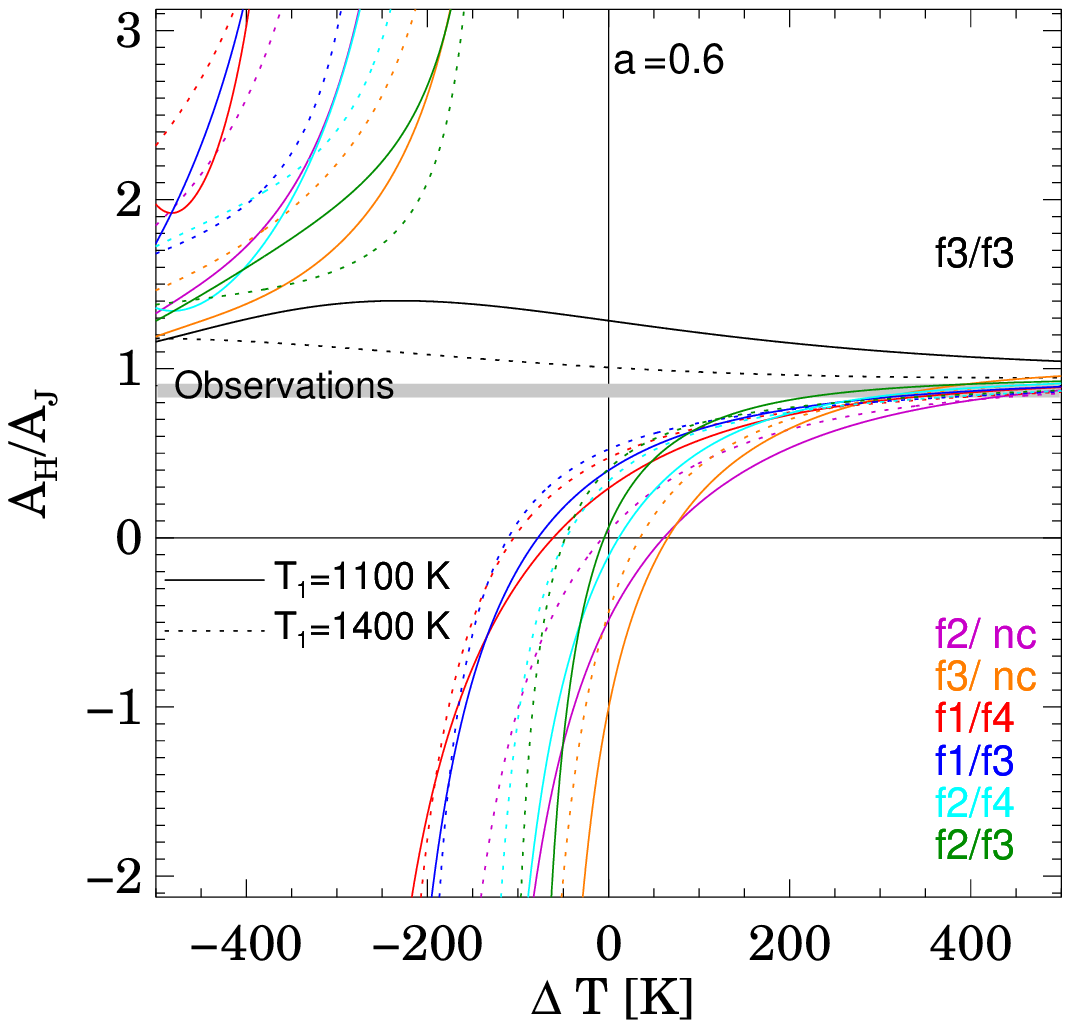}
\end{array}$
\vspace{0cm}
\caption{{\small {\em Top:} Model amplitude ratios $A_{K_s}/A_J$ for a BD with a heterogeneous surface as a function of the temperature contrast between surface elements, $\Delta T$.  Amplitude ratios were computed from equation \ref{eq:var2}, using a variety of cloudy and clear model fluxes ($\mathcal{F}_1[f_{\rm sed1},T_1]$ and $\mathcal{F}_2[f_{\rm sed2},T_2+\Delta T$]) from \citet{saumon08}, indicated by color in the plot legend.  Here we have used the shorthand of $fn_1/fn_2$ to denote $f_{\rm sed1}=n_1/f_{\rm sed2}=n_2$, and `nc' for the cloud-free model.  For all cases curves are plotted for constant $T_1$=1100~K (solid lines) and $T_1$=1400~K (dotted lines).  Colored lines correspond to case $A$ (heterogeneous clouds) while the black lines correspond to case $B$ (cool/hot spots).  The left and right panels demonstrate a weak dependence on the initial filling fraction of the $\mathcal{F}_2$ regions, with $a$=0.2 (left) and $a$=0.6 (right).
{\em Middle:} For each curve in the above panel, the corresponding change in filling factor, $\Delta a$, required to produce the maximum observed variability for \targ of $A_J=$0.26, is shown. {\em Bottom:} Same as the top panels but for $A_{H}/A_J$ ($\Delta a_{26}$ remains unchanged from the upper panel).}
\label{fig:mod1a}}
\end{figure*}

\colb{For each pairing of model grids specified by $f_{\rm sed1}$=\{1,2,3\} and $f_{\rm sed2}=\{3,4,$ nc$\}$ } we used equation \ref{eq:var2} to compute (i)$A_{K_s}/A_{J}$ and $A_{H}/A_J$ as a function of $\Delta T$, and (ii)the corresponding change in filling factor, $\Delta a$ required to produce the maximum observed variability of $A_J$=0.26.  \colj{Results for a representative selection of these pairings} are shown in figure \ref{fig:mod1a} for values of $T_1$=1100~K and 1400~K; a reasonable range for 2M2139.  \colj{ Results are plotted for both $a=0.2$ and $a=0.6$ in order to demonstrate a weak dependence of our result on the minimum filling factor.  Note that the overall effective temperature can be much greater than $T_1$ when $\Delta T > 0$, as warm regions quickly dominate the total flux.  Although it is not clear a priori what appropriate values of $T_1$ should be, 1400~K represents a reasonable upper limit when $\Delta T>0$, since it coincides with both our model fitting in section \ref{sect:single} and the upper limit of the L/T transition temperature found by \citet{stephens09}.  The lower value of $T_1$=1100~K is shown to be appropriate in the following section. }

\begin{figure}
\hspace{-0.8cm}
\epsscale{1}
\plotone{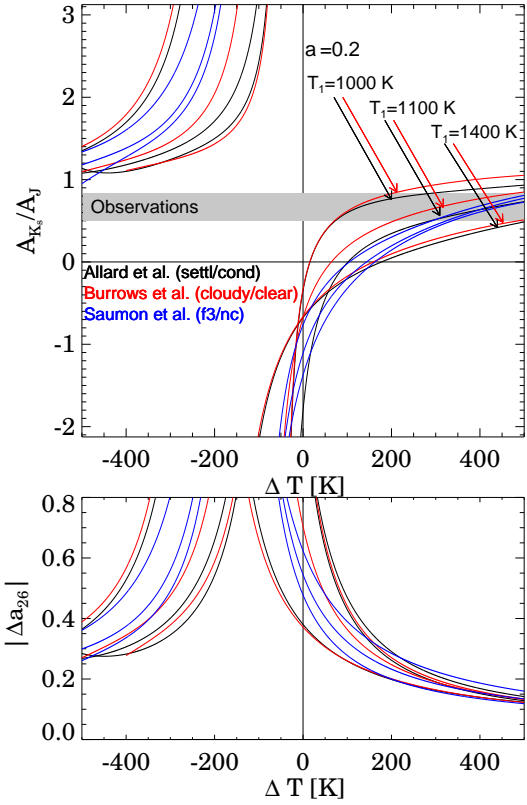}
\caption{Same as figure \ref{fig:mod1a}, with $a=0.2$, and including comparisons to the cloudy/clear and SETTL/COND models of \citet{burrows06} and  \citet{allard01,allard03} (see the main text for more detail).   For each set of models the three curves correspond to fixed values of $T_1$=1100~K, 1200~K, and 1400~K.  
\label{fig:mod2}}
\end{figure}

For comparison, figure \ref{fig:mod2} compares results for case $A$ from different model atmosphere groups, including the cloudy and clear models of \citet{burrows06} and the SETTL and COND models of \citet{allard01,allard03}.

Our figures \ref{fig:mod1a}-\ref{fig:mod2} recover trends previously discussed by \citet{artigau09}.  Namely, cool or hot spots within a cloudy atmosphere (case $B$) produce amplitude ratios $A_{K_s}/A_{J}$ exclusively $>1$, whereas warm regions of low-condensate opacity (case $A$) yield $0\lesssim A_{K_s}/A_{J}\lesssim1$ for $\Delta T\gtrsim100$~K .   \vvnew{The observed amplitude ratios for \targ are positive and consistently  $<$1 and therefore inconsistent with the presence of magnetic spots.  Rather, our modeling supports an interpretation of patchy clouds, wherein regions of lower condensate opacity are warmer.}  

\colj{As seen in figures \ref{fig:mod1a}-\ref{fig:mod2}, a variety of model combinations for case $A$ are able to reproduce observations, with differing requirements for $\Delta T$ and $\Delta a$.   In general, models where the difference between  $f_{\rm sed1}$ and $f_{\rm sed2}$ is small require lower temperature contrasts but sightly larger changes in filling factor to model the observations.  Increasing the parameter $a$ yields large amplitude ratios for a given temperature contrast $\Delta T$, but also requires a larger $\Delta a$.  In all cases temperature contrasts $\gtrsim 150~K$ are required to model the observed amplitude ratios, with some model combinations requiring temperature contrasts in excess of 400 K.  The SETTL/COND and cloudy/clear models of \citet{allard03} and \citet{burrows06} are equally able to reproduce the photometric variations, requiring somewhat lower temperature contrasts when $T_1\lesssim1200$~K.  While the variety of models and parameters capable of reproducing the photometric data appears large, in the following section we use \targs's NIR spectrum as an additional constraint to significantly reduce the span of good solutions.}

\subsection{A specific variability model for \targ}

\begin{figure*}
\centering
\includegraphics[width=5.5in]{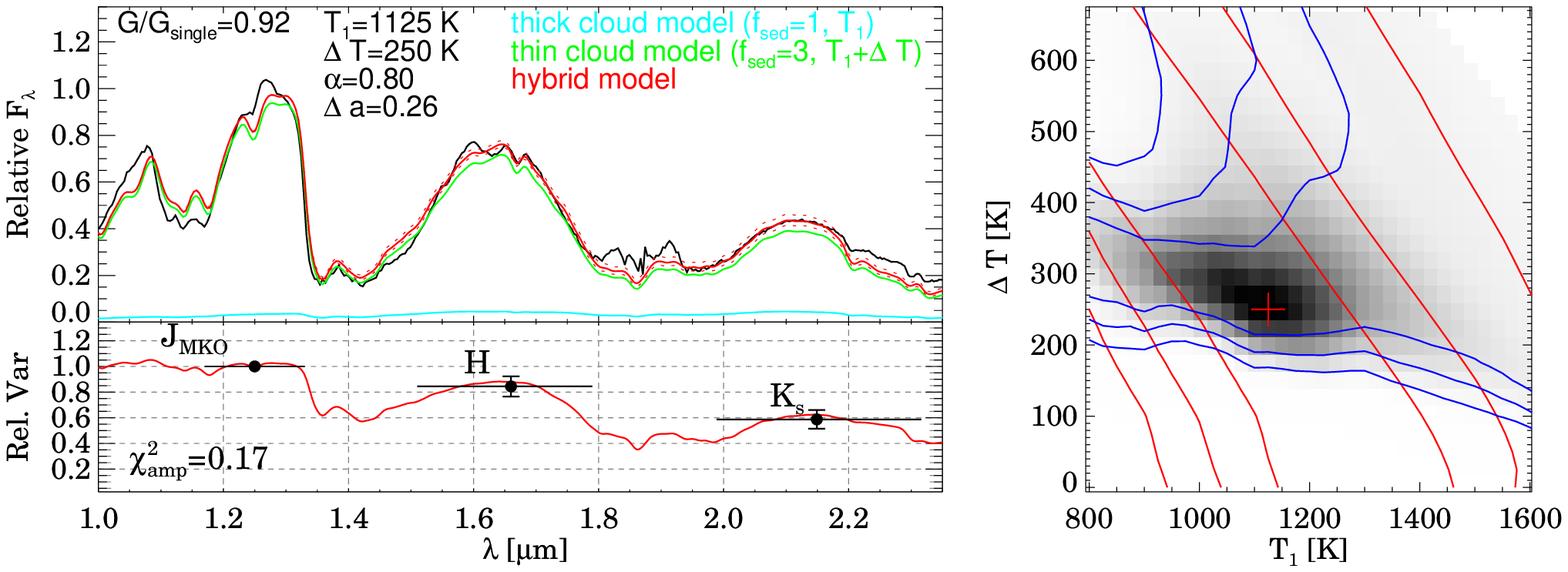} \\[-3mm]
\includegraphics[width=5.5in]{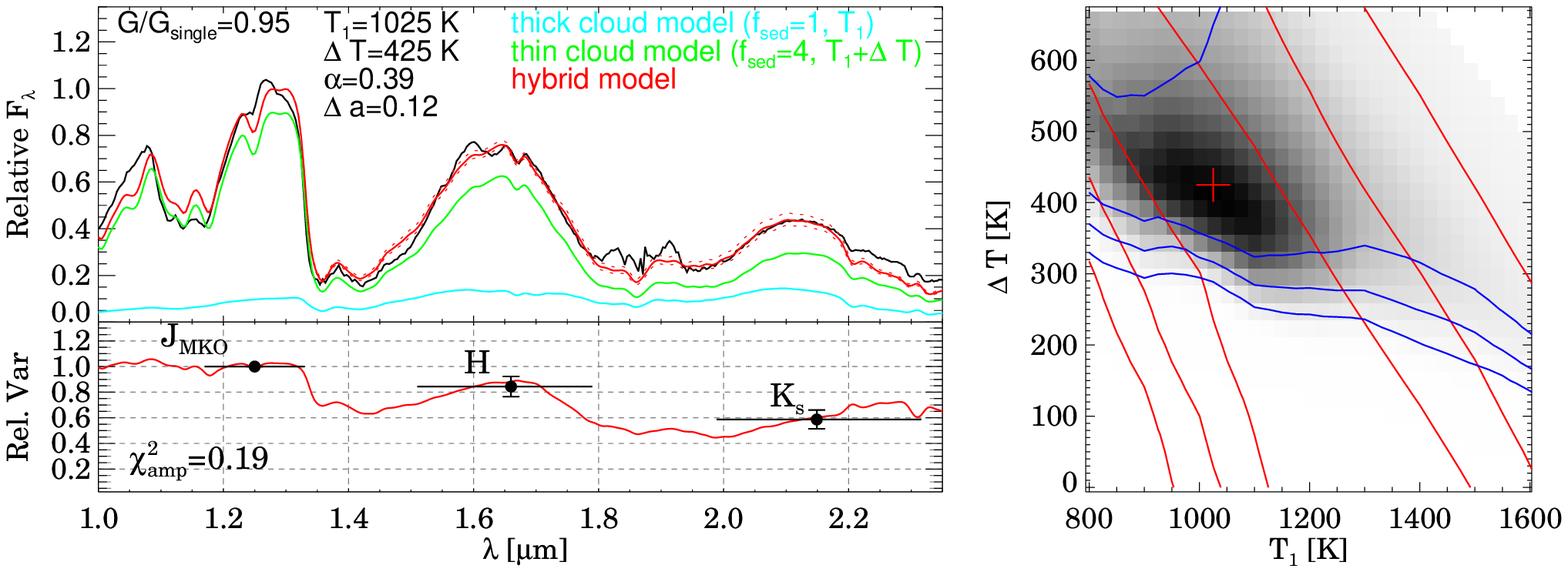} \\[-3mm]
 \includegraphics[width=5.5in]{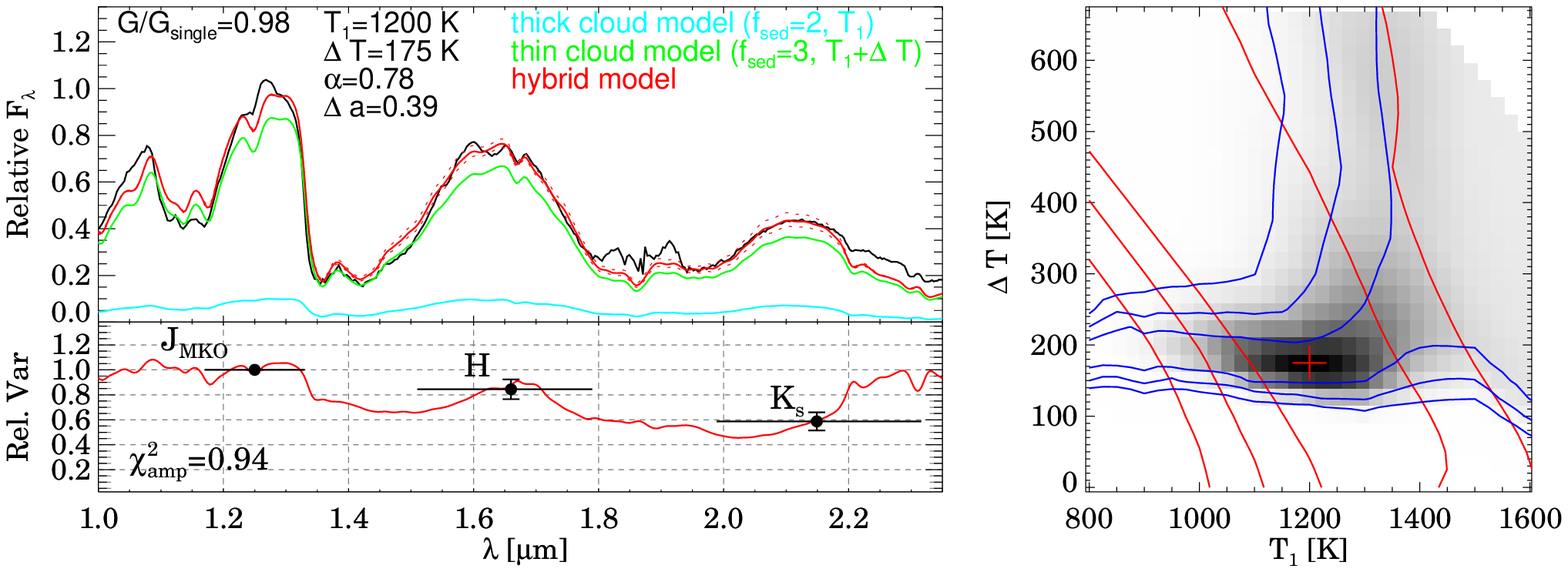} \\[-3mm]
  \includegraphics[width=5.5in]{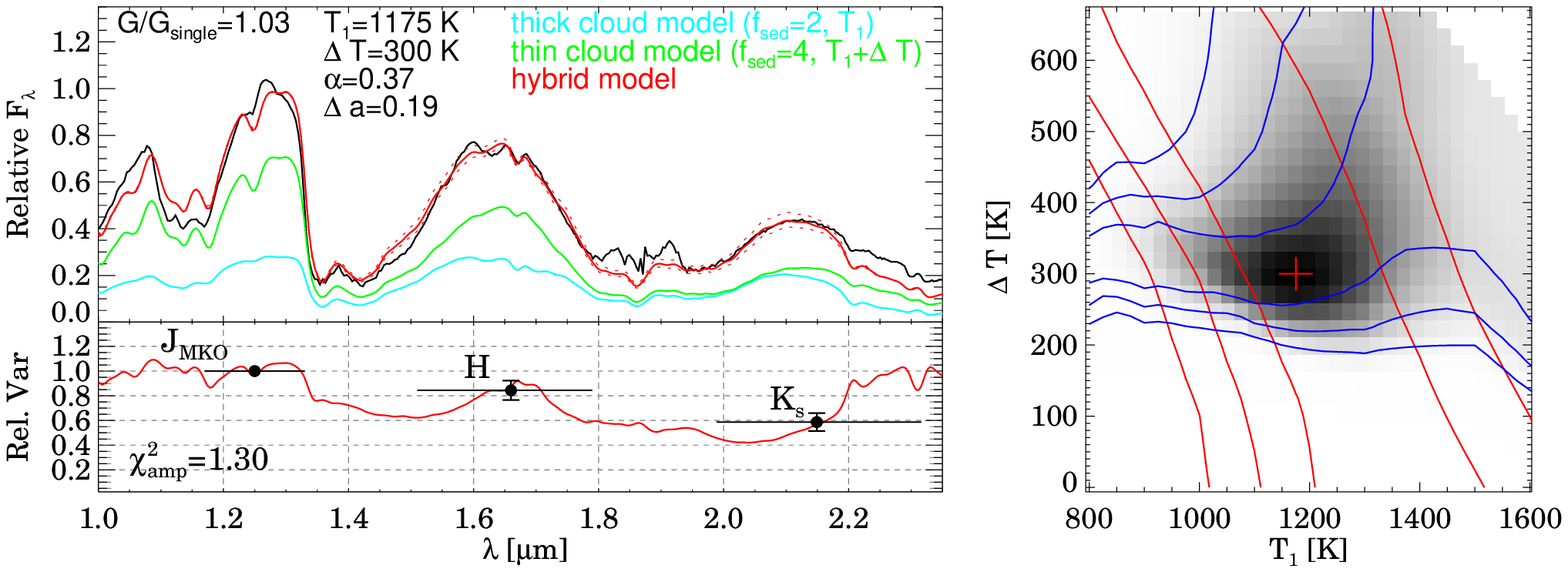}
  \vspace{0mm}
  \caption{\colb{Simultaneous fits to both the NIR spectrum (black line) and photometric variability (black data points) of \targ for the four model combinations where $f_{\rm sed1}=\{1,2\}$ and $f_{\rm sed2}=\{3,4\}$ (see text for explanation). {\em Right:} Contours in the $T_1$-$\Delta T$ plane for $\Delta \chi^2_{\rm amp}$ (blue) and for \postsub{$4\, \Delta G/G_{single}$} (red), corresponding to levels of \postsub{$\{2.3,6.2,11.8\}$}, where the $\Delta$ indicates differences in $G$ and $\chi^2_{amp}$ from their minimum values on the grid.   
  A red cross marks the combination of $T_1$ and $\Delta T$ (for which $\alpha$ and $\Delta a$ are also uniquely determined) \postsub{where $\chi^2_{\rm amp}$+$4\, G/G_{\rm single}$ is a minimum.}   
  \postsub{The shaded map shows the exponential of $-(\chi^2_{\rm amp}+4\, G/G_{\rm single})/2$ with square-root scaling to emphasize the tails of the distribution.}
{\em Left:} The hybrid model spectrum at mid-light (red solid line), corresponding to the best-fit parameters marked by the red cross on the right.  The model is a linear combination of cloudy (cyan line) and less-cloudy (green line) surfaces, where the filling fraction of less-cloudy regions varies by an amount $\Delta a$.  Dotted red lines show the hybrid model at maximum and minimum light, assuming $A_J$=0.26, and normalized over the 1.23 to 1.25 $\mu$m range. The bottom left panel shows the relative variations that would result for the best-fiting model as a function of wavelength, normalized to the $J$ band.  The observed amplitude ratios and their errors are over plotted for comparison, and $\chi^2_{\rm amp}$ for the best fit is specified.}}
 \label{fig:contours}
\end{figure*} 

\label{sect:mod}
 The modeling of photometric variations alone, as done in the previous section,  ignores additional information about \targs's spectral energy distribution that can further constrain the nature of the variability. In addition, it has been previously noted \citep[e.g.][]{burrows06}, that the same models capable of reproducing observed changes in broadband colors do not necessarily reproduce the NIR spectra of L/T transition dwarfs.   
 
 For combinations of $f_{\rm sed1}=\{1,2,3\}$ and $f_{\rm sed2}=\{3,4,{\rm nc}\}$ and $\log{g}=4.5$, we searched for combinations of $T_1$, $\Delta T$, $a$, and $\Delta a$ capable of reproducing {\em both the NIR SED and observed variations} of 2M2139.  We have restricted ourselves to the region of parameter space where
 $\Delta T>0$ corresponding to the upper right quadrant in figure \ref{fig:mod1a}.  
 We systematically examined a grid of hybrid models with $T_1=800-1600$~K and $\Delta T=0-700$~K in steps of 25~K.
 Our warmest model has $T_{eff}=2100$~K, and therefore our grid is incomplete for regions where $T_1+\Delta T>2100$~K.  We note that all original spectra were computed with 100~K spacing, and we have linearly interpolated between models to obtain our finer sampling.  At each grid point:
 
\begin{enumerate}
\item A parameter $\alpha$ (corresponding to the mid-brightness filling factor, $a$+0.5$\Delta a$) was uniquely determined by fitting a hybrid model given by 

\begin{align}
M(\lambda)&=
(1-\alpha)\mathcal{F}_1[\lambda;T_1,f_{\rm sed1}]
\nonumber
\\[1mm]
&\qquad\qquad\quad+\alpha \mathcal{F}_2[\lambda;T_1+\Delta T,f_{\rm sed2}]
\label{eq:hyb}
\end{align}

to \targ's NIR spectrum, using a procedure identical to the one used for spectral fitting in section \ref{sect:single} (see equation \ref{eq:fit}).   We note that in the above equation we have used $\mathcal{F}_1$ and $\mathcal{F}_2$ to represent model fluxes as a function of wavelength, instead of integrated over a given bandpass as before.  The minimum weighted $\chi^2$ statistic, $G$, associated with each grid point was recorded.

\item We solved for $a$ and $\Delta a$ by inputing $a=\alpha-0.5\Delta a$ and $A_J$=0.26 into equation \ref{eq:var2}, where we have chosen to set $A_J$ to the maximum observed variability.\footnote{\colj{When $\alpha$ is fixed, $\Delta a$ is linear in $A_J$, and therefore the $\Delta a$ corresponding to smaller overall amplitudes can be easily inferred.}}  
In some cases $\Delta a$ derived from the best spectral fit was unphysical (e.g. $a+\Delta a > 1$), and in these cases we systematically varied our fit parameter $\alpha$ until we found the lowest-$G$ fit that also satisfied the physical condition that $a+\Delta a \le 1$.

\item  We computed model amplitude ratios $A_{K_s}/A_J$ and $A_H/A_J$ by inputting parameters $a$, $\mathcal{F}_1[T_1]$, and $\mathcal{F}_2[T_1+\Delta T]$ into equation \ref{eq:var2}.  A $\chi^2$ statistic for the amplitude ratios given by \postsub{$\chi^2_{\rm amp}=[(A_H/A_J - x)^2/\sigma_x^2+(A_{K_s}/A_{J}-y)^2/\sigma_y^2]$} was recorded, where $x$ and $y$ are the observed amplitude ratios, and $\sigma_x$ and $\sigma_y$ their uncertainties (see table \ref{tab:red}).
\end{enumerate}

According to the above procedure (repeated separately for various combinations of $f_{\rm sed1}$ and $f_{\rm sed2}$), $G$ and $\chi^2_{\rm amp}$ were determined for every grid point in the $T_1$-$\Delta T$ plane, allowing us to identify regions of parameter space capable of reproducing both the NIR spectrum and broadband variability of 2M2139.  

\postsub{We obtained best-fit parameters by determining where $\chi_{\rm amp}^2$+4\,$G/G_{\rm single}$ was a minimum on our grid.
Here, $G_{\rm single}$ corresponds to the best-fitting single (non-hybrid) spectral model found in section \ref{sect:single}.   We note that the $G$ values have $N-3$ degrees of freedom (dof; where $N=304$ is the number of data points used for fitting) and have been scaled such that $G_{\rm single} =$ dof.  If we were to minimize $\chi_{\rm amp}^2$+$G$, the best-fit parameters would be strongly dominated by the spectral fit, which uses significantly more data points (however, due to model incompleteness and highly correlated residuals this increase in data points does not result in proportionately more precise parameters).  
Instead, we have used our prior knowledge of the observed spectrum to impose a more broad constraint, allowing a wide range of ``credible'' spectral templates and eliminating obvious mismatches.  In practice, we found that scaling $G$ by $4\, /G_{\rm single}$ assigns a reasonable weight to the spectral fit (e.g., see figures \ref{fig:contours} and \ref{fig:simp0136}).}

We found only a few combinations of $f_{\rm sed1}$ and $f_{\rm sed2}$ that were capable of reproducing the photometric variations, while at the same time providing a reasonable spectral match.   Those where cool, high condensate opacity regions are represented by $f_{\rm sed1}=\{1,2\}$ and warm, lower condensate opacity regions are represented by $f_{\rm sed2}=\{3,4\}$ generally provide good fits to the data, with the $f_{\rm sed1}=1$/$f_{\rm sed2}=3$  combination providing the best fit.  These four model scenarios, \vnew{and corresponding best-fit parameters}, are shown in figure \ref{fig:contours}, \colb{fitted to the simultaneous $JHK_s$ photometric data from 30 Sep 2009}.  
\postsub{Shown are contours for $4\,G/G_{single}$ and $\chi_{\rm amp}^2$ in the $T_1$-$\Delta T$ plane.  }
The hybrid model spectrum corresponding to the best-fit parameters, as well \colb{as the resultant variability as a function of wavelength are also shown}.  \colb{We have chosen to show fits to the $A_{H}/A_J$ and $A_{K_s}/A_J$ amplitude ratios from 30 Sep 2009 as the amplitude ratios measured on this date are approximately intermediate to those measured in the other two epochs for which we have simultaneous $JHK_s$ light curves.  We have opted not to average the amplitude ratios over all epochs since the measurements are incompatible at the $2 \sigma$ level and may represent different cloud configurations.   In particular, the near-unity amplitude ratios from 26 Sep 2009 are challenging to fit, and strongly bias the averaged values.  Rather, we provide best-fit contours and spectra corresponding to the other two epochs  in Appendix \ref{sect:ap2}.}

\vnew{Depending on the combination of $f_{\rm sed1}$ and $f_{\rm sed2}$  we infer temperature contrasts of $\Delta T$=175-425\s K between thick and thin cloud regions, and changes in filling factor of $\Delta a$=0.12-0.39 (see figure \ref{fig:contours}). }

 Rather than an atmosphere composed of clouds and clearings, our observations are best reproduced by one possessing regions of differing cloud opacity.  For the \colb{best-fitting pairing of $f_{\rm sed1}=1/f_{\rm sed2}=3$ shown in figure \ref{fig:contours}  (and similarly for the $f_{\rm sed1}={2}/f_{\rm sed2}={3}$ combination) the data can be described by an atmosphere composed primarily of $f_{\rm sed2}=3$ regions (67\% -\postsub{93}\% coverage over the visible disc as a function of phase) with cooler, thicker cloud patches ($f_{\rm sed1}=1$) covering the remainder of the surface (\postsub{7}\%-33\%).  The thick cloud patches have a temperature of $T_1$=\postsub{1125~K}, while the thinner cloud patches are $\Delta T=250~K$ warmer.  Physically, this model could indicate condensate clouds that form in at least two distinct layers at different depths in the atmosphere.  Such layers could arise due to compositional stratification of condensates \citep[e.g.][]{lodders06}, or from complex atmospheric dynamics.   Long-lived cyclonic storm systems and transient high altitude clouds are seen in the atmospheres of the Solar System gas giants \postsub{ \citep[e.g.][]{smith89,vasavada98}}, and it is possible that similar features exist in the atmosphere of 2M2139.  
Since small spatial scales will tend to be averaged out in the disc-integrated light, the large amplitude of our observations (and the corresponding large azimuthal asymmetry in modeled cloud coverage) is most consistent with the presence of sizable cloud features.  If the observed variations were attributed to a single high-altitude storm, it would have to occupy \postsub{26\%} of the visible disc in order to produce $A_J=$0.26.}
 
\colb{Alternatively, for the next-best-fitting $f_{\rm sed1}={1}$/$f_{\rm sed2}={4}$ model also shown in figure \ref{fig:contours} (and similarly for the $f_{\rm sed1}={2}/f_{\rm sed2}={4}$ combination) the data can be described by an atmosphere composed primarily of cloudy $f_{\rm sed1}=$1 regions with $T_1$=\postsub{1025\s K} (67\%-55\% coverage as a function of time) interspersed with warmer regions of lower condensate opacity ($f_{\rm sed2}=4$).  In this case a much higher temperature contrast of $\Delta T=425~K$, and a correspondingly smaller change in filling factor affecting only 12\% of the visible disc are required.  This scenario is closer to the picture of an atmosphere covered by cloudy and clear regions that might naively be expected based on theories of cloud fragmentation at the L/T transition.  However, it is notable that we are unable to find solutions where $f_{\rm sed2}$=`nc' to represent the data, indicating that the modeled ``clearings'' must retain some degree of condensate opacity (i.e. $f_{\rm sed2}=4$).}
\vnew{
For an atmosphere with a same T-P profile throughout and where temperature decreases monotonically as a function of altitude (as is expected for isolated BDs), this large temperature contrast is an indication that thick cloud patches extended to higher, cooler regions of the atmosphere.
}

\colb{For the best-fitting hybrid spectra, the normalizing constant $c=(R/d)^2$ in Eq. \ref{eq:fit},  along with a BD radius of $R$=0.13 R$_{\odot}$ obtained from the evolution models of \citet{saumon08} yield spectroscopic distances ranging from 15.7-18.5\s pc, which are larger than the 11.6\s pc distance derived earlier from the empirical $K$-band relationship of \citet{marocco10}.  If we instead assume a higher gravity of $\log{g}=5.0$ at odds with the best-fitting spectral models, we find $R=0.1$\s  R$_{\odot}$ and obtain distances of 13.8-14.1\s pc, which are closer to the empirical estimate.}

We also attempted the above analysis using the SETTL/COND and cloudy/clear model combinations of \citet{allard03} and \citet{burrows06}, however we were unsuccessful in simultaneously reproducing the NIR SED and photometric variations of \targ using these models.  This is not surprising given that we find no good solutions with $f_{\rm sed2}=$~`nc' (truly condensate-free regions) using the models of \citet{saumon08}.

\colb{Finally, we should note that in finding the best hybrid models, we have assumed that our target's observed spectrum corresponds to its mid-brightness state.  This is a reasonable approximation, as synthetic colors derived from 2M2139's NIR spectrum are consistent with $J-K_s$ and $J-H$ colors measured for this target at the time of the WIRC observations.  In figure \ref{fig:contours} we display the model maximum and minimum light spectra in addition to the best-fitting mid-brightness spectrum, all normalized to the $J$ band, in order to demonstrate that differences between the three are minimal.   Even when the absolute variability amplitudes are large and differ as a function of wavelength ($A_J$=0.26, $A_{K_s}/A_J=0.45$-$0.83$), changes to the shape of a normalized spectrum are barely perceptible by eye.  Accordingly, the assumption that the observed spectrum corresponds to mid-brightness does not significantly impact our results.  In contrast, the obvious spectral mismatch that results from fitting models to photometric variations alone, without using a spectral constraint, is illustrated in figure \ref{fig:simp0136} for SIMP 0136.}

\subsection{Comparison to SIMP0136}
 \begin{figure*}
\centering
\includegraphics[width=3in]{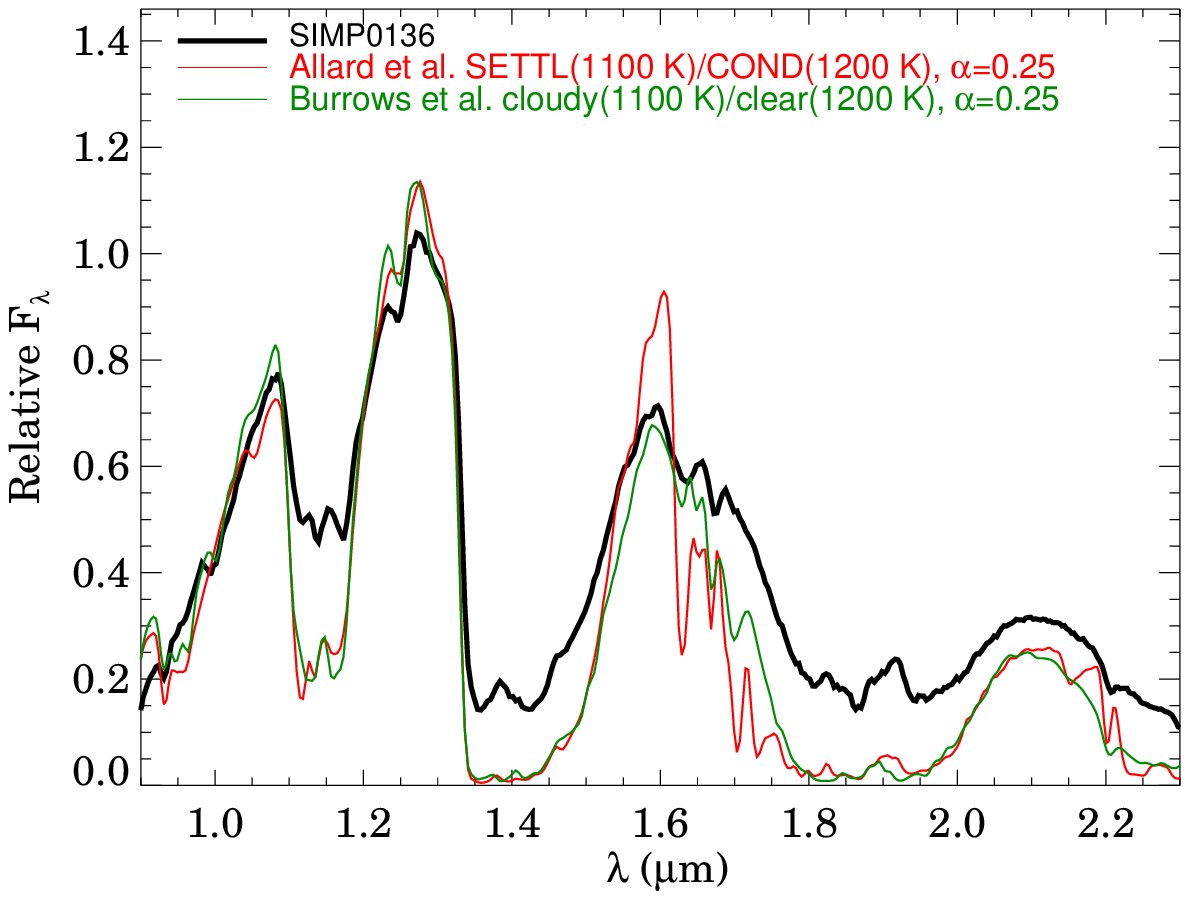} 
\includegraphics[width=3in]{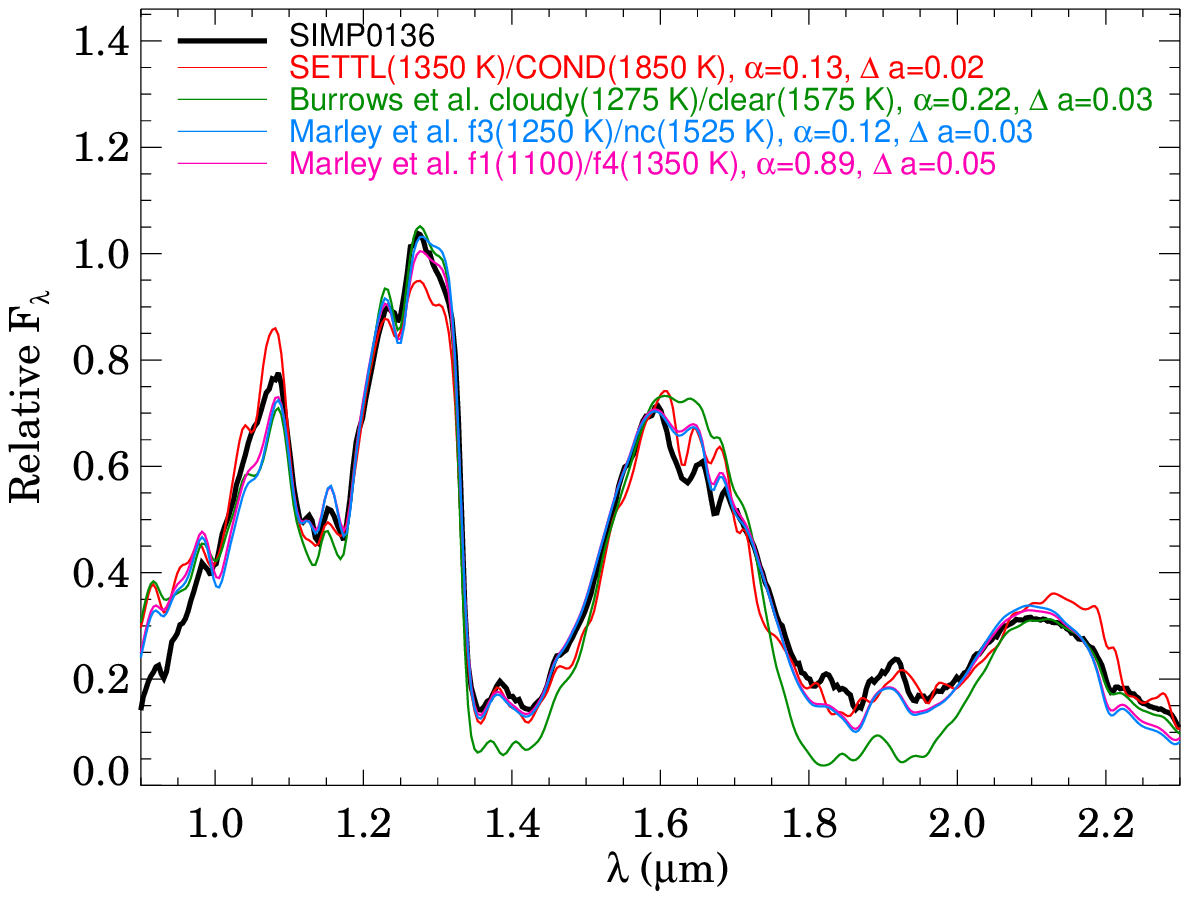} 
 \caption{The NIR spectrum of SIMP0136 \citep{burgasser06}, with hybrid model spectra (linear combinations of cloudy and clear models) used to model its photometric variability overplotted (similar to the left panel of figure \ref{fig:contours}).  The filling fractions and effective temperatures of the models used to form each hybrid spectrum are shown in the legend.  {\em Left:} Hybrid model spectra obtained by setting $T_1+\Delta T$=1200~K a priori,  and modeling only photometric variations as done by \citet{artigau09} (see text for more detail).
{\em Right:} Hybrid model spectra corresponding to our simultaneous modeling of the photometric variations {\em and} NIR spectrum of SIMP0136.  Red and green lines correspond to the same model sets as in the left panel, but with filling fractions and temperature contrasts that better reproduce SIMP0136's NIR spectrum as well as its photometric variability of $A_{K_s}/A_J=0.48$.  Blue and pink lines correspond to models of \citet{saumon08} where  \colb{$f_{\rm sed1}=1$/$f_{\rm sed2}=4$  and $f_{\rm sed1}=f4$/$f_{\rm sed2}=$`nc' respectively.}
 \label{fig:simp0136}}
 \end{figure*}

Here we apply our variability model to SIMP0136, the other known example of a quasi-periodic variable T-dwarf. \citet{artigau09} modeled the variations of SIMP0136 ($A_{K_s}/A_J\sim 0.48$ and $A_J\sim0.05$~mag) using 1D cloudy and clear model atmospheres and an interpolation scheme equivalent to our equation \ref{eq:var2}.  While we have used primarily the models of \citet{saumon08}, \citet{artigau09} used combinations of cloudy and clear models from 3 different groups: 
the SETTL and COND models of \citet{allard01,allard03} and the cloudy and clear models of \citet{burrows06}, as well as those from \citet{tsuji03}.  They found that a temperature contrast of $\Delta T\sim100$~K between clear and cloudy regions and $\Delta a\sim$0.15 can reproduce the $J$ and $K_s$ variability of SIMP0136, with good agreement between all model sets.  However, we find that the model spectra corresponding to this result ($T_1$=1100~K, $\Delta T$=100~K, $a$=0.25) provide a poor match to SIMP0136's NIR spectrum (figure  \ref{fig:simp0136}), demonstrating that models capable of reproducing changes in broadband colors do not necessarily yield good spectral matches.

In order to make a fair comparison between \targ and SIMP0136 we have repeated our above analysis for SIMP0136, relying on the amplitudes of variability reported by \citet{artigau09} and SIMP0136's NIR spectrum from \citet{burgasser06}.  We find that the variability and NIR spectrum of SIMP0136 can be well-reproduced by a wide range models, including those where $f_{\rm sed1}=\{1,2,3\}$ and $f_{\rm sed2}=$4, as well as those where $f_{\rm sed1}=\{3,4\}$ and $f_{\rm sed2}=$~nc.  These solutions require temperature contrasts ranging from $\sim$75~K for the $f_{\rm sed1}=3$/$f_{\rm sed2}=$4 pairing to $\sim$275~K for the $f_{\rm sed1}=3$/$f_{\rm sed2}=$nc pairing, and require changes in filling factor affecting 3\% to 15\% of the visible disc.  We also repeated our analysis using the SETTL/COND models of \citet{allard03} and cloudy/clear models of \citet{burrows06}.  For these latter models we found poor overlap between parameters capable of simultaneously reproducing the observed photometric variations and SIMP0136's NIR spectrum, with optimal temperature contrasts occurring for $\Delta T$=500~K and 300~K respectively, and significantly poorer spectral fits.  A selection of best-fitting hybrid spectra for SIMP0136 from different model groups is shown in figure \ref{fig:simp0136}.   

In general, we find that our results agree qualitatively with those of \citet{artigau09}.  However, for the sake of comparison, we find that temperature contrasts required to model the observed $A_{K_s}/A_J\sim 0.48$ for SIMP~0136 are at least 2-3 times higher (and changes in filling factor smaller) than previously inferred if a spectral constraint is imposed.

\subsection{The need for self-consistent 3D model atmospheres}
\label{sect:cons}
Due to the lack of self-consistent 3D models our above modeling approach is necessarily data-driven as opposed to physically motivated.  The drawback is that combinations of independent 1D atmosphere models may possess incompatible T-P profiles, and we are therefore not guaranteed to arrive at physically plausible configurations \citep[see the discussion within][]{marley10}.  This leads us to question whether the high temperature contrasts derived above, and in particular $\Delta T \gtrsim$~300~K in several cases, are physically reasonable.  

\begin{figure}
\includegraphics[width=2.7in,angle=-90]{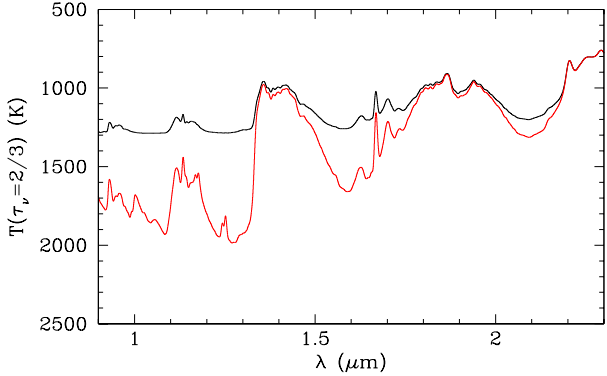} 
\caption{Temperature of the $\tau=2/3$ surface as a function of wavelength for a model with $f_{\rm sed}=2$, $\log{g}=4.5$, and $T_{eff}$=1000~K, with (black line) and without (red line) condensate opacity, using models of \citet{saumon08}. \label{fig:tbright}}
\end{figure}

Figure \ref{fig:tbright} shows the temperature of the $\tau=2/3$ surface as a function of wavelength for a model with $f_{\rm sed}=2$, $\log{g}=4.5$ and $T_{eff}$=1000~K, with and without condensate opacity. 
 Both models share the same T-P profile, but in the latter condensate opacity has been switched off.  The cloud-free $\tau=2/3$ surface occurs at higher temperatures, indicating that photons emerge from deeper atmospheric layers in the absence of condensate opacity.  The temperature contrast between the cloudy and cloud-free ``photospheres'' is as high as 600~K in the $J$ band, and therefore the large temperature contrasts required to model our observations are not unreasonable.  However, the relative contrasts in the $J$ and $K_s$ bands appear problematic; in the $K_s$ band the contrast between the cloudy and cloud-free $\tau=2/3$ surface has dropped to only $\sim$100~K.  Thus, in the $K_s$ band, one does not look much deeper into the atmosphere even if cloud opacity is removed.   This is mainly due to an increased gaseous opacity from collision induced absorption by $H_2$ molecules (CIA) at these wavelengths, causing the gaseous (cloud-free) photosphere to occur at higher altitudes, and hence cooler temperatures.   This structure of the cloudy model poses a problem if we wish to reproduce the large observed amplitude ratios of $A_{K_s}/A_J\gtrsim 0.45$, as it requires the $K_s$ band flux within clear regions to be larger than allowed by the T-P profile of the cloudy model.   Of course, this assumes that the T-P structure of the upper atmosphere is identical within cloudy and clear regions, which need not be the case.

The above results suggest that conclusions drawn from linear combinations of 1D atmosphere models must be viewed with caution, and highlights the need for self-consistent 3D modeling, including the effects of atmospheric circulation.  Recent progress in this direction has been made by \citet{marley10}, who have calculated a set of self-consistent patchy cloud models of the L/T transition, where cloudy and clear regions share a common T-P profile.  These models continue to use the cloud model of \citet{ackerman01} to describe cloudy regions and contain an additional parameter $h$, which specifies the fractional coverage of ``holes'', or cloud-free regions in the atmosphere.  However, these models predict negative $A_{K_s}/A_J$ (i.e. anti-correlated amplitudes), inconsistent with those observed for \targs.  \colb{Since the \citet{marley10} models assume a common T-P profile within cloudy and clear regions, this may indicate that the T-P profiles diverge in a real atmosphere.  Alternatively, our modeling in the previous section demonstrates that the best fits to the data derive from models in which the lowest condensate opacity surface elements are not truly condensate-free (i.e. $f_{\rm sed}=4$), which may also account for part of the discrepancy between the data and models from \citet{marley10}.}

There are several possible explanations for the mismatch between self-consistent models and the $K_s$ band data.
A larger than expected contrast between cloudy and clear regions in the $K_s$ band could suggest thick, high-altitude dust clouds, or weaker than expected H$_2$ CIA, possibly the result of high metallicity. 
\colb{Alternatively, the T-P structure of a partially cloudy atmosphere may itself be horizontally inhomogeneous within the photosphere, requiring a fully 3D hydrodynamical treatment to model.}
In addition, there is always the possibility that different types of cloud heterogeneities than envisioned here, or a completely different mechanism may  be responsible for the variability of \targ.   Assuming that the variability is indeed the result of heterogeneous clouds, these observation can be used to guide future modeling efforts.  

\section{Discussion}
\label{sect:discuss}

\subsection{Origin of the Observed Variability}
\colj{While heterogeneous clouds is our preferred explanation for the observed variability, and the scenario we have modeled in depth, it is worth exploring whether other mechanisms could be responsible.} 
\subsection{Clouds versus Magnetic Spots}
The asymmetric and evolving light curve shape of 2M2139, with periodic behavior on a rotation timescale, is the typical signature of surface spots.   As noted earlier, the derived 7.721$\pm$0.005~hr periodicity would be consistent with rotation rates of ultracool dwarfs \citep[$\sim$2-12~hrs][]{reiners08} inferred from $v\sin{i}$ data, whereas a double-peaked light curve with a 15.44~hr period would correspond to a slower rotator \colb{(this is not inconsistent with $v\sin{i}$ data, as observations are biased against such slow rotators).}  The nature of the surface features responsible is not immediately clear.  For stars, cool magnetic spots are a common source of variability, related to the suppression of convective heat flux within magnetic flux tubes that penetrate the stellar photosphere.  However, estimates of magnetic Reynolds numbers in cool dwarf atmospheres \citep{gelino02,mohanty02}, predict weak coupling between the gas and magnetic field, with magnetic Reynolds numbers of $R_m \sim$ 0.01-$10^{-6}$ throughout the photospheres of late-L and T type BDs.  It is unknown to what extent cooling in deeper, ionized layers might carry through to the photosphere.  From an empirical standpoint, only a small fraction of late-M dwarfs display variability due to magnetic spots, and typical photometric amplitudes are $< 5$\% \citep[e.g.][]{tendrup99,scholz04,scholz09}.   Thus, if magnetic in origin, the NIR variability of \targ would be an order of magnitude larger than what is typical of late M-dwarfs.

Alternatively, condensate clouds in rapidly rotating BD atmospheres may form discrete cloud features, similar to the banding and cyclonic storms observed on Jupiter.  At wavelengths where the gaseous opacity is low (e.g. in the NIR 1-2 $\mu$m regime between water absorption bands), the flux emitted from `holes' or clearings in the cloud layer may emerge from atmospheric depths hundreds of degrees hotter than the opaque cloud layer (e.g. figure \ref{fig:tbright}).  Such large temperature contrasts would provide a natural explanation for large-amplitude variability in the $J$ and $H$ bands, and somewhat smaller variability in $K_s$.  On Jupiter, regions of low condensate opacity are visible as bright hot spots at 5 $\mu$m, where an analogous gaseous opacity window exists.  By conducting synthetic photometry on resolved images of Jupiter, \citet{gelino00} determined that the contrast between cloud features and hot spots in Jupiter's atmosphere may lead to variability as high as 20\% at 5$\mu$m over a single rotation, with variations strongly correlated with the position of  the Great Red Spot, a high altitude anti-cyclone.  Thus Jupiter may prove to be the closest astrophysical analog for \targs's unique variability, with both warm clearings and a high-altitude storm feature contributing to variations.

\colj{Finally, the modeling presented here (\S \ref{sect:model}, figures \ref{fig:mod1a}-\ref{fig:mod2}) strongly supports the heterogeneous clouds interpretation --- rather than magnetically induced cool or hot spots --- as it reproduced the observations best.   Since magnetic structures and clouds would be effectively decoupled, the magnetic spot model would correspond to the case B we have investigated (uniformly cloudy atmosphere with spatial variations in $T_{eff}$ only), which we found cannot explain the observations.  This is of course consistent with the fact that clouds are known to occur in BD atmospheres while magnetic spots are very unlikely in these cool atmospheres.}  

At this point we cannot distinguish between a scenario where a high altitude cloud or storm feature is responsible over a scenario where the atmosphere is composed primarily of broken clouds and clearings.  Further monitoring, and improved  atmosphere models where heterogeneities are represented in a more physically consistent way will be required to confirm this result. 
 
\subsubsection{Could the variability of \targ be related to binarity?}
\begin{figure}
\epsscale{1.1}
\plotone{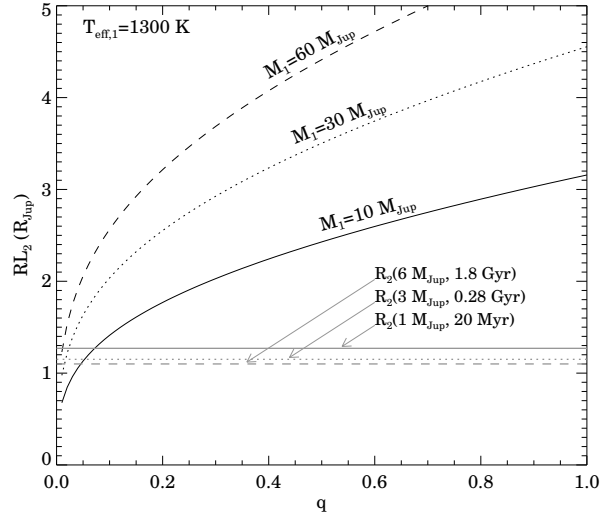}
\caption{Secondary Roche Lobe radius as a function of mass ratio, $q$, for primary masses of 10, 30, and 60 $M_{Jup}$, assuming an orbital period of 15.44~hr (black lines).  Secondary radii determined from evolutionary models for fixed $q=0.1$, and assuming a primary effective temperature of 1300~K are overplotted (grey lines).
\label{fig:bin}}
\end{figure}

Given \targs's potentially double-peaked light curve, and similar amplitudes of variations observed in $J$, $H$, and $K_s$ bands in at least one epoch (see figure \ref{fig:dblpk}) we consider here whether the observed variability could arise from an eclipsing binary system.  
Since the continuous shape of the light curve is inconsistent with that of a detached binary system (there is no flat out-of-transit region), a semi-detached 
configuration would be required.  In addition, subtle changes in the light curve shape from night to night, as well as the overall light curve asymmetry would require an extra variable component, such as a hot-spot due to mass transfer.  The requirement that one component fill its Roche Lobe allows us to place constraints on the range of primary and secondary masses and ages that could form a semi-detached system.  
For a given primary mass, $M_1$, and mass ratio $q=M_2/M_1$ (where $M_2<M_1$), the Roche Lobe size of the secondary can be computed via the formula of \citet{eggleton83} given by $R_L/a=0.49 q^{-2/3}/(0.6 q^{-2/3}+\ln{[1+q^{-1/3}]})$.  The separation, $a$, is determined from Kepler's law assuming a period of 15.44~hr, corresponding to a double-peaked light curve.  The secondary Roche Lobe sizes can then be compared to typical BD radii from evolutionary models.  In figure \ref{fig:bin} we show the secondary Roche Lobe size as a function of mass ratio for primary masses of 10, 30 and 60 $M_{Jupiter}$.  For comparison we over-plot typical secondary BD radii from evolutionary models, assuming a primary effective temperature of $\sim$1300~K (note that the system age is fixed from the combination of primary mass and effective temperature, yielding approximate ages of 20~Myr, 280~Myr, and 1.8~Gyr from lowest to highest primary mass).  The system parameters for which secondary Roche Lobe filling may occur are limited.  We find that Roche Lobe sizes are much larger than typical secondary radii for almost all primary masses and mass ratios. This scenario is possible only for low primary masses ($M_1<10$~$M_{Jup}$) and low mass ratios ($q<0.05-0.1$), such that a semi-detached binary with a 15.44~hr period would have to be a young ($<$ 20 Myr), double planetary-mass system.  For more typical BD masses and ages, the binary components should remain detached, and unable to produce the observed light curve.

\colb{Phase variations (i.e. from the day/night side contrast of an irradiated companion) in non-eclipsing systems can also produce periodic variations.  However, intrinsic phase variations of an irradiated companion would be washed out by the brighter primary, resulting in much lower levels of variability than observed for \targs.}

A further problem with the binary scenario is that multi-epoch observations spanning Aug 2009 to Nov 2009 cannot be phased together, requiring either the Aug 2009
or Nov 2009 epoch to be offset in phase from the other epochs.  The timing of both epochs has been verified in the acquisition time-stamps as well as manual log books, and phasing all epochs together would require both to be in error---an occurrence we deem to be extremely unlikely.  In order to produce a large offset in the expected eclipse timings by natural means, a moderately elliptical, precessing orbit is required ($e\gtrsim 0.15$, e.g. \citet{ragozzine09}).  This could potentially be achieved if the system is undergoing Kozai oscillations \citep{kozai62} due to the presence of a third body in a highly inclined orbit to the inner pair.  However, in order for the precession timescale to be sufficiently fast (e.g. months) such a third perturber would closely approach or exceed the criterion for dynamical instability given by \citet{eggleton95}.  Furthermore, the effects of tidal friction between the inner pair would act to suppress Kozai cycles over time and circularize the inner orbit \citep[e.g.][]{fabrycky07}, making it increasingly unlikely to observe a system in this configuration.

Based on the above considerations we feel we can reasonably exclude an eclipsing binary as the source of \targs's variability.  

Alternatively, like Jupiter, BDs are expected to have large magnetic fields. Thus it is conceivable that some sort of magnetic interaction between a close binary pair is possible without requiring physical Roche lobe filling.  However, such a scenario does not naturally explain a double-peaked light curve, and it's ability to produce significant and highly contrasting photospheric features is unclear.

\subsection{The L/T Transition}
\begin{figure}
\hspace{-8mm}
\epsscale{1.2}
\plotone{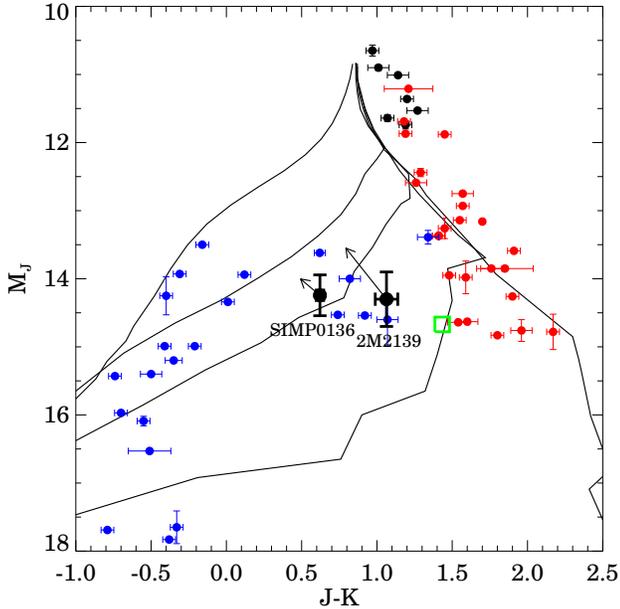}
\caption{Color-magnitude diagram in the MKO filter system for M, L, and T dwarfs with measured parallaxes (black, blue and red points respectively) from the compilation of S. K. Leggett.
 The approximate positions of SIMP0136 and \targ are overplotted based on spectroscopic parallaxes as opposed to real ones.  The $J-K_s$ color of \targ (converted to the MKO system) is based on the colors derived from WIRC and SpeX epochs (see table \ref{tab:phot}), as opposed to 2MASS.  The approximate position of \targ on the color magnitude diagram based on its 2MASS color (also converted to the MKO system) is shown as a green square (its absolute magnitude was determined {\em relative to} the WIRC/SpeX epochs, as the empirical relations for determining absolute magnitude have not changed).  Solid black lines from right to left indicate the $f_{\rm sed}=$1,2,3,4, and nc model colors respectively for $\log{g}=$5.0.  \colj{Although we have used spectral models with $\log{g}=4.5$ throughout, a value of $\log{g}=5.0$ is more appropriate for comparison with the population of field BDs (in addition, see the discussion in section \ref{sect:single} about choices of $\log{g}$ values).}
Arrows depict the vector direction of the variability observed for \targ (this paper, averaged from the 3 epochs in figure \ref{fig:ijhk}) and reported for SIMP~0136 \citep{artigau09}, both magnified in amplitude by a factor of 3.
\label{fig:cmd}
}
\end{figure}

\targ joins SIMP0136 as the second known cool BD to display {\em persistent and significant} quasi-periodic variability in the NIR, with amplitudes well above the level of photon noise and systematic errors.
With spectral types of T1.5 and T2.5, and NIR colors of $J-K_s=1.3$ and $J-K_s$=0.98, both BDs fall squarely within the L/T transition regime (see figure \ref{fig:cmd}).  Although a larger sample is needed, observations of \targ and SIMP0136 suggest that large amplitude NIR variability may be more common at the L/T transition, consistent with the cloud fragmentation hypothesis of \citet{ackerman01} and \citet{burgasser02_lt}.  In addition, the positions of SIMP0136 and 2M2139 on the color-magnitude diagram --- within the L/T transition regime where the cloud opacity decreases significantly --- provides further rationale to prefer heterogeneous clouds as the mechanism for the observed variability.

\colb{
Overplotted on the color magnitude diagram in figure \ref{fig:cmd} are vectors showing the direction of observed variability for \targ and SIMP~0136.  Rather than tracing the approximate evolution of field BDs across the L/T transition, the vectors point upward of the transition path, rising much more steeply in $M_J$ with decreasing $J-K$.  This difference reflects the fact that evolution across the L/T transition occurs over astronomical timescales, over which a significant amount of heat is lost from the upper atmosphere as the higher temperature T-P profile below a cloudy atmosphere evolves toward the lower temperatures under a clear atmosphere \citep[e.g.][]{saumon08}.   When we observe variability due to clouds and clearings at a snapshot in time we see underlying clear patches that are warmer than the integrated effective temperature of the atmosphere, and hence warmer than the atmosphere of a more evolved cloud-free T-dwarf of the same effective temperature.  Therefore, the direction of the L/T transition path is determined by both decreasing cloud coverage and a loss of entropy from the upper photosphere that occurs over astronomical timescales, while the direction of the instantaneous variability vector traces only changes in cloud coverage. }

While heterogeneous clouds appear to be the most likely explanation for our observations, the physical mechanism by which clouds may fragment at the L/T transition remains unknown.  One possibility is that as clouds form progressively lower in the atmosphere as the BD cools, eventually the entire vertical extent of the cloud is found within the dynamic troposphere which is more subject to local variations in updrafts and downdrafts arising from convection.  These localized weather patterns are better able to disrupt the cloud than when the BD is warmer and a substantial fraction of the cloud opacity is found within the relatively quiescent stratosphere.   While weather remains a poorly understood phenomenon even among the planets in our own Solar System, BDs represent a simplified case where atmospheric dynamics result primarily as a consequence of rapid rotation and internal heat, without the complication of external forcing due to irradiation from a parent star.  Thus future observations of weather in BD atmospheres may provide a novel opportunity to study atmospheric circulation and cloud meteorology in a higher gravity regime, never before probed.  

\section{Conclusions}
\label{sect:concl}
We have reported the unique, large-amplitude quasi-periodic variability of the cool L/T transition BD 2M2139.  Our major findings and conclusions are summarized here.

~\\
{\em 1. Variability}  We have observed \targ to be highly variable in the NIR over 9 separate epochs spanning 100 days, with an amplitude as high as 26\% in the $J$-band.  Measured amplitude ratios range from $A_J/A_H=0.84-0.91$ and $A_{K_s}/A_J=0.45-0.83$, with  uncertainties of $\pm$0.07-0.15.  It remains unclear whether these variations in relative amplitude from epoch to epoch are intrinsic, or reflect hidden measurement uncertainties. Disagreement at the 2$\sigma$ level between epochs suggests the former.  In addition to the short timescale variability, comparison with 2MASS photometry hints that there may be a longer-term (10~yr) variation that consists mainly of a brightening in the J-band.\\

~\\
{\em 2. Periodicity} We have measured a minimum period of 7.721$\pm$0.005~hr for 2M2139.  It is also possible that the light curve is double-peaked with a period of $2\times 7.72=15.44$~hr. In the latter case, the period would be somewhat larger than the 2-12~hr periods typical of field BDs.  

~\\
{\em 3. Time Evolution of the Light Curve} Despite a clear periodicity, the light curve shape changes from night to night, and we cannot simultaneously phase our earliest and latest epochs with the middle ones, indicating substantial evolution of cloud features on timescales of weeks to months, with more subtle changes occurring on a timescale of days.  The differential rotation of a storm feature is a possible explanation for the longer timescale evolution observed.

~\\
{\em 4. Modeling variability due to patchy clouds}  We have searched for linear combinations of 1D atmosphere models differing in temperature and condensate properties, that can simultaneously reproduce the observed multi-band variability and the NIR spectrum of \targs.   Using the models of \citet{saumon08} we found that the observations can be reproduced by a heterogeneous surface \colb{wherein regions of higher condensate opacity ($f_{\rm sed}$=1 or $f_{\rm sed}$=2) are cooler, and regions of lower condensate opacity ($f_{\rm sed}=3$ or $f_{\rm sed}=4$) are warmer (by $\sim$175-425~K).  The best-fitting model suggests that cool, thick cloud features--- for instance one or more storms systems occupying 25\% of \targs's visible disc ---may be responsible.  Alternatively, the data is also well-reproduced by a model where atmospheric heterogeneities consist of warm clearings in a cloudy atmosphere, however ``clear" regions must retain some degree of cloud opacity (i.e. $f_{\rm sed}$=4).  The large temperature contrast required between thin and thick cloud patches may suggest that the thick clouds extend to higher altitudes in the atmosphere.}  We reiterate that these results should be viewed cautiously, as interpolations between independent 1D models do not necessarily result in physically plausible configurations.

~\\
{\em 5. Origin of Variability}  The continuous, asymmetric, and evolving nature of the light curve strongly suggests that atmospheric cloud features are responsible for \targs's variability.  Due to \targ's (i)status as an L/T transition BD and the expectation of patchy clouds in this regime, (ii)cool, neutral photosphere, and (iii)our modeling of the observations
we have argued that surface features responsible are heterogeneous clouds rather than magnetically induced spots.  
We have also considered that \targ may be an interacting binary, but have ruled out this scenario.  

~\\
{\em 6. The L/T Transition}  \colb{Both \targ and SIMP0136, the two known examples of BDs that display large-amplitude, persistent variability in the NIR, have colors and spectral types falling directly within the L/T transition, suggesting that variability in this regime may be higher-amplitude and/or more frequent.  These results therefore provide empirical support for the idea that the fragmentation of dust clouds can explain observed properties of the L/T transition.}

Looking forward, long term monitoring of this target both photometrically and spectroscopically, and over a broader wavelength regime, should reveal the true nature of its variability.  In addition, a parallax measurement and high resolution spectrum will be essential in order to better constrain \targ's absolute magnitude and physical properties.

\acknowledgments
This work was supported in large part by Research Tools \& Instrumentation and Discovery grants, a Steacie Fellowship, and the Canada Research Chairs program, all from the Natural Sciences and Engineering Research Council, to RJ.  JR is supported in part by a Vanier Canada Graduate Scholarship from the National Sciences and Engineering Research Council of Canada.  Work by DS was supported in part by two Spitzer Science Center grants and by the United States Department of Energy under contract DE-AC52-06NA25396. We thank Ian Thompson of Carnegie Observatories and the staff of the Las Campanas Observatory for their help in scheduling and carrying out the observations. This research has benefitted from the SpeX Prism Spectral Libraries, maintained by Adam Burgasser at http://pono.ucsd.edu/~adam/browndwarfs/spexprism.  This publication makes use of data products from the Two Micron All Sky Survey, which is a joint project of the University of Massachusetts and the Infrared Processing and Analysis Center/California Institute of Technology, funded by the National Aeronautics and Space Administration and the National Science Foundation.

\appendix
\section{A. Relative Flux Calibration of SpeX Prism Library Spectra and Synthetic 2MASS Colors}
\label{sect:ap1}
The relative flux calibration between $J$, $H$, and $K_s$ bands for M, L, and T dwarf sources in the SpeX Prism Library was investigated by determining synthetic 2MASS colors from the SpeX spectra and comparing them to reported values from the 2MASS catalog.  Synthetic 2MASS colors were found using the relative spectral response curves and zero-magnitude fluxes provided by \citet{cohen03}.  The error estimates, $\sigma_{SpeX}$ are the sum of a random error component determined from the measurement uncertainties associated with individual wavelength bins (typically small), and an additional 0.07 mag corresponding to the average difference in synthetic $J-K_s$ colors computed for (presumed non-variable) targets which have been observed at two epochs within the SpeX Prism Libraries.  

\begin{figure*}[here]
\centering
\includegraphics[width=3.33in]{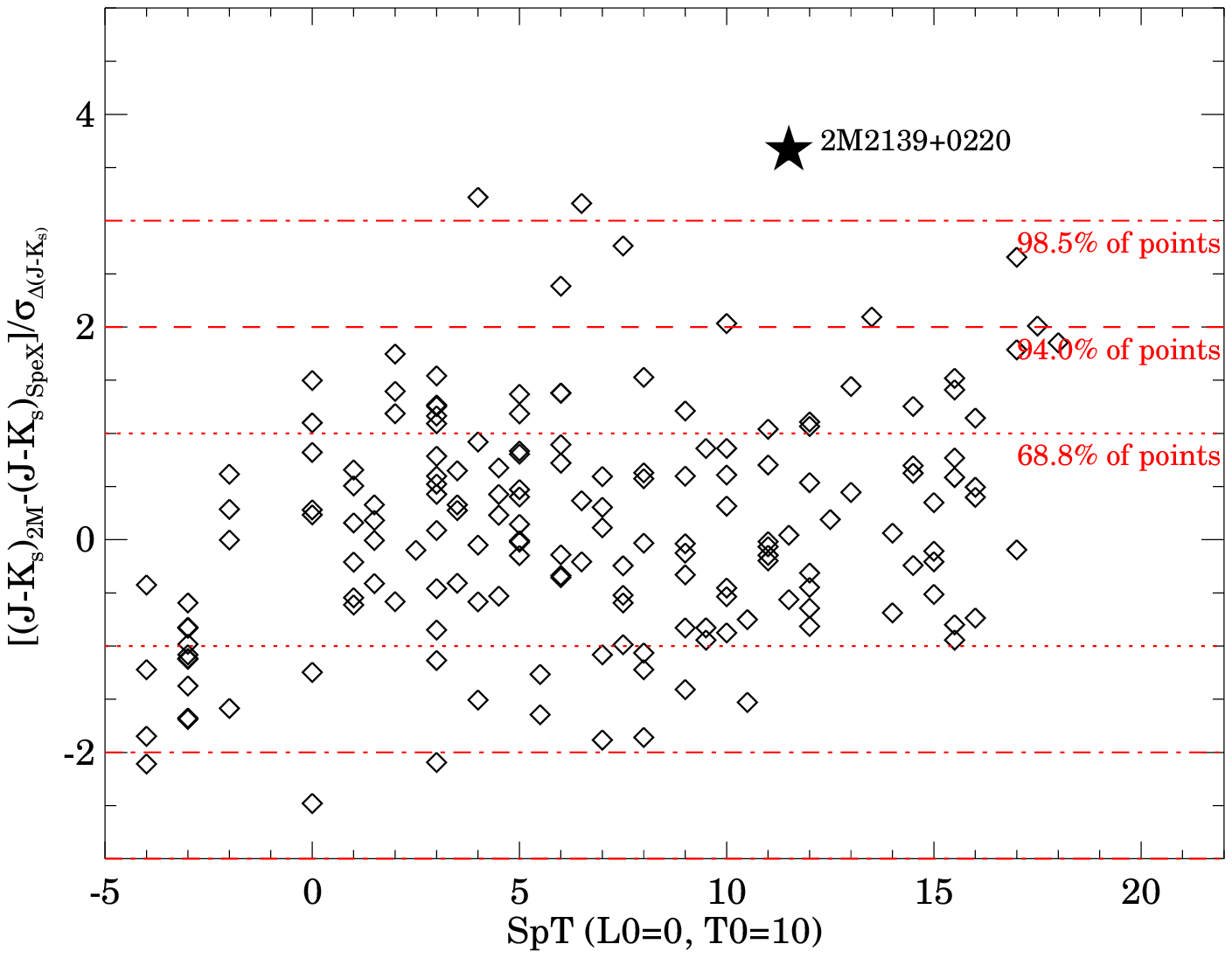}
 \includegraphics[width=2.7in]{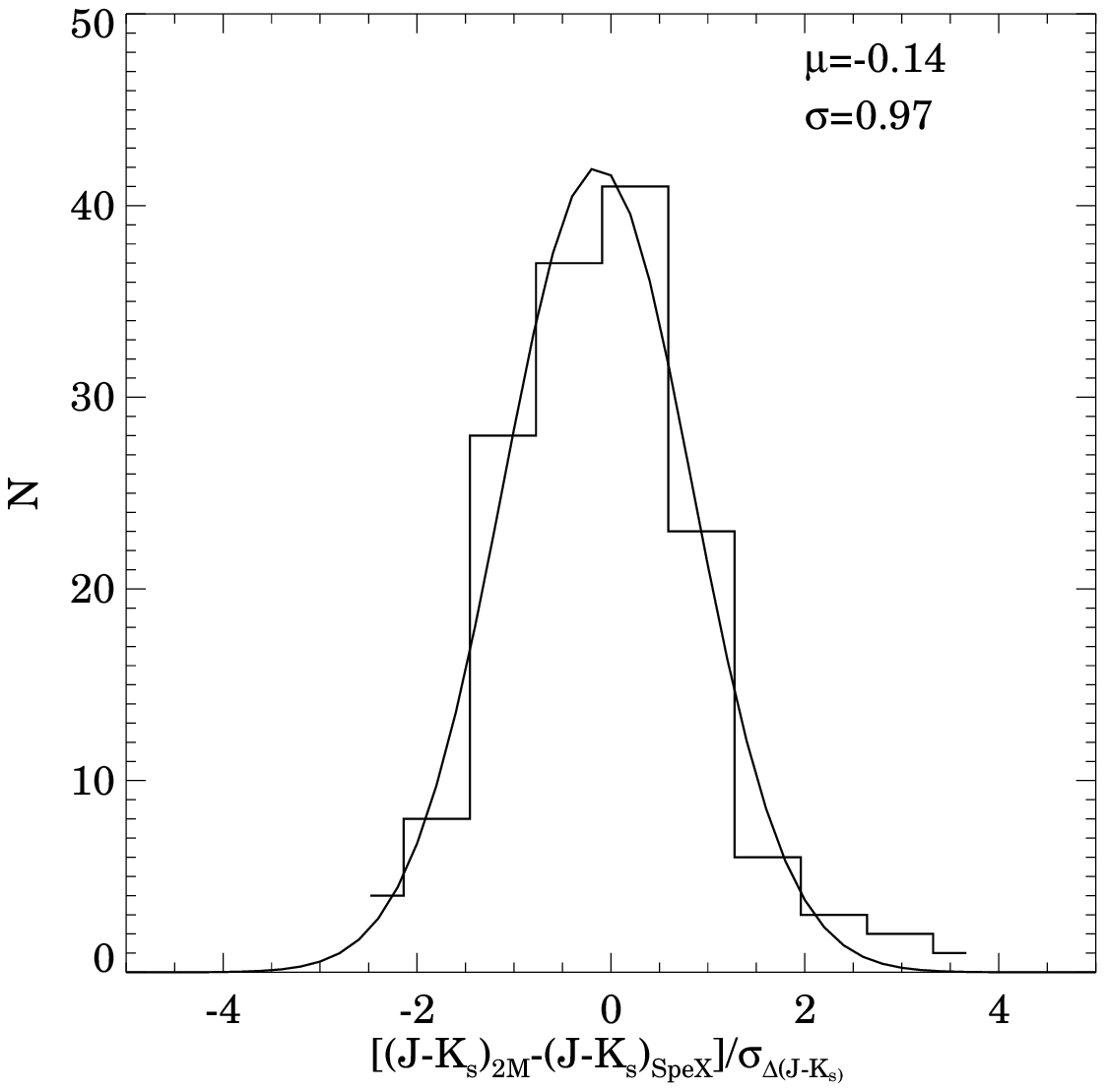}
\caption{Comparison between 2MASS photometry and synthetic 2MASS colors derived from SpeX prism spectra for all L and T dwarfs in the SpeX prism library.  Sources flagged in the prism library as having a poor SNR, or those with poor quality $J$ and/or $K_s$ 2MASS photometry (Point Source Catalog quality flag not equal to `A',`B',or `C'), were omitted. \label{fig:spex}}
\end{figure*}

Differences between 2MASS colors and SpeX synthetic colors were determined for all M, L and T dwarf sources from the SpeX Prism Library that also had moderate to high quality detections in 2MASS (2MASS catalog QFLG of at least `C').  For each BD meeting these requirements, the difference between the 2MASS and synthetic SpeX colors, $\Delta(J-K_s)$, was computed and assigned an uncertainty of

\begin{equation}
\sigma_{\Delta(J-K_s)}=\sqrt{\sigma_J^2+{\sigma_{K_s}^2+\sigma_{SpeX}^2}}
\end{equation}

where $\sigma_J$ and $\sigma_K$ are the photometric errors taken directly from the 2MASS catalog.  The resultant $\Delta(J-Ks)$/$\sigma_{J-K_s}$ for all BDs considered is plotted in figure \ref{fig:spex}, and shows that \targ is a 4$\sigma$ outlier.  If we consider only L and T-dwarfs the match between 2MASS and SpeX colors is surprisingly good; a gaussian fit to a histogram of $\Delta(J-Ks)$/$\sigma_{J-K_s}$, also shown in figure \ref{fig:spex}, has width of $\sigma\sim$1 and is roughly centered about zero ($\mu$=-0.14). While differences in $J-K_s$ color are roughly symmetric about zero for L and T dwarfs, there appears to be a small systematic offset for M dwarfs, the origin of which is not clear.  

Therefore, uncertainties in synthetic $J-K_s$ colors for SpeX Prism Library L and T dwarf spectra are well estimated by propagating the standard errors reported for each wavelength bin, and then adding to this 0.07~mag.

\section{B. Models for \targs's Variability at Additional Epochs}
\label{sect:ap2}
\begin{figure}
\centering
\includegraphics[width=5.5in]{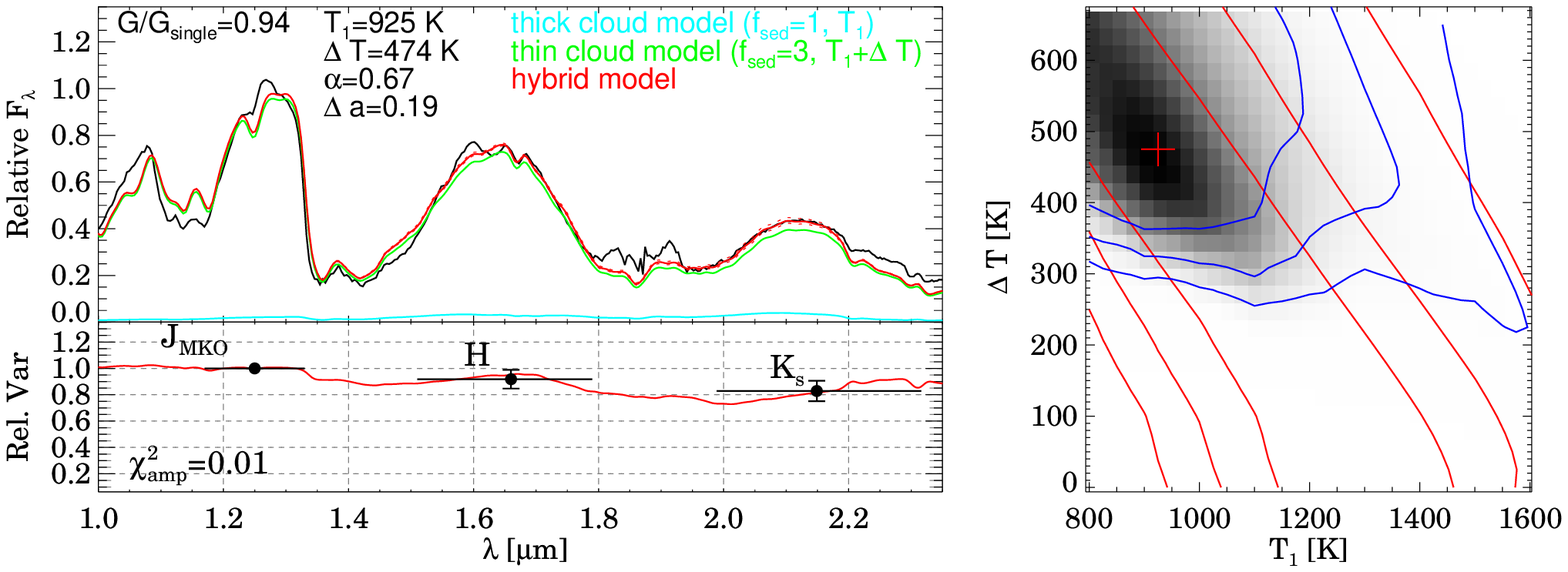} \\[-3mm]
\includegraphics[width=5.5in]{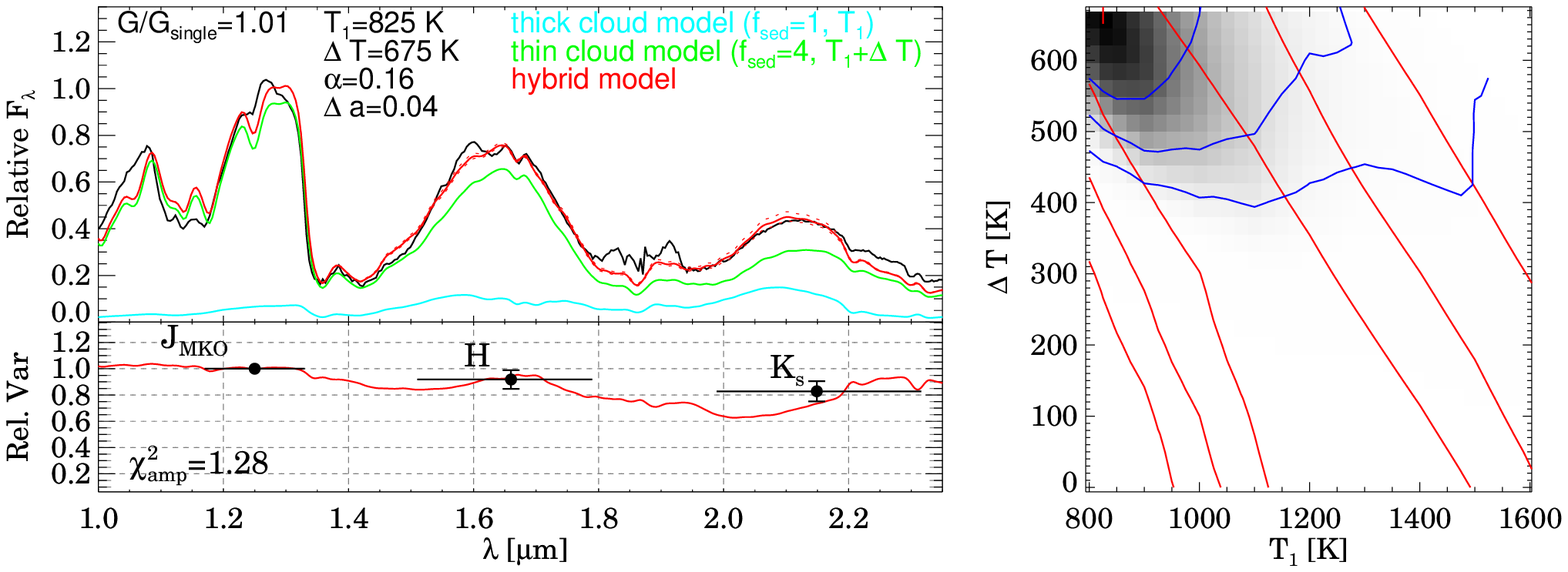} \\[-3mm]
 \includegraphics[width=5.5in]{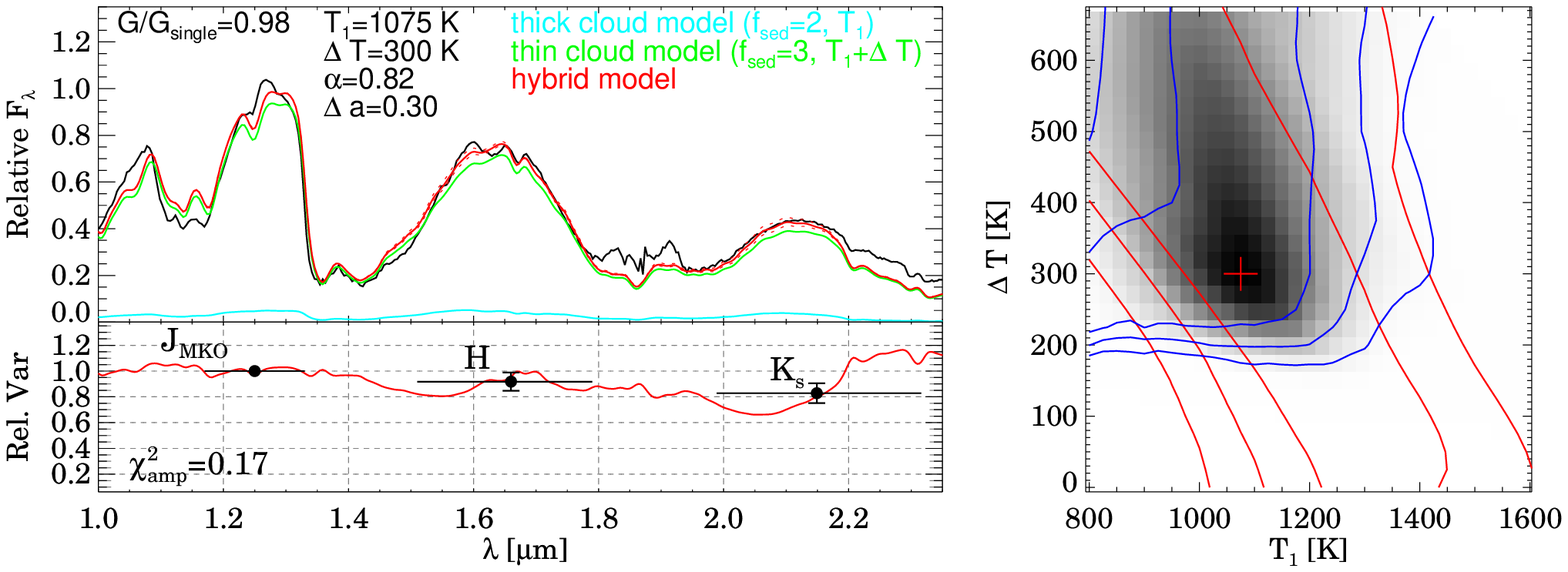} \\[-3mm]
  \includegraphics[width=5.5in]{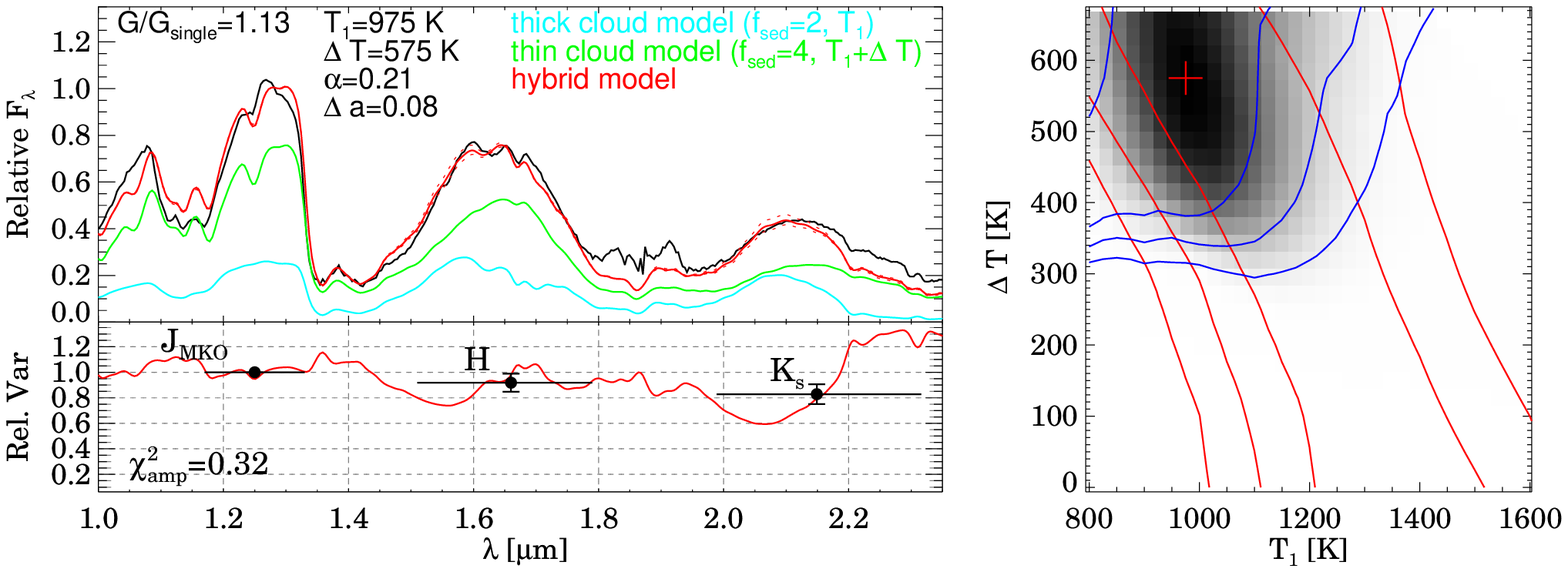}
  \vspace{0mm}
\caption{Same as figure \ref{fig:contours}, but with amplitude ratios constrained using data from the 26 Sep 2009 epoch.}
\label{fig:extra_mod1}
\end{figure}

\begin{figure}
\centering
\includegraphics[width=5.5in]{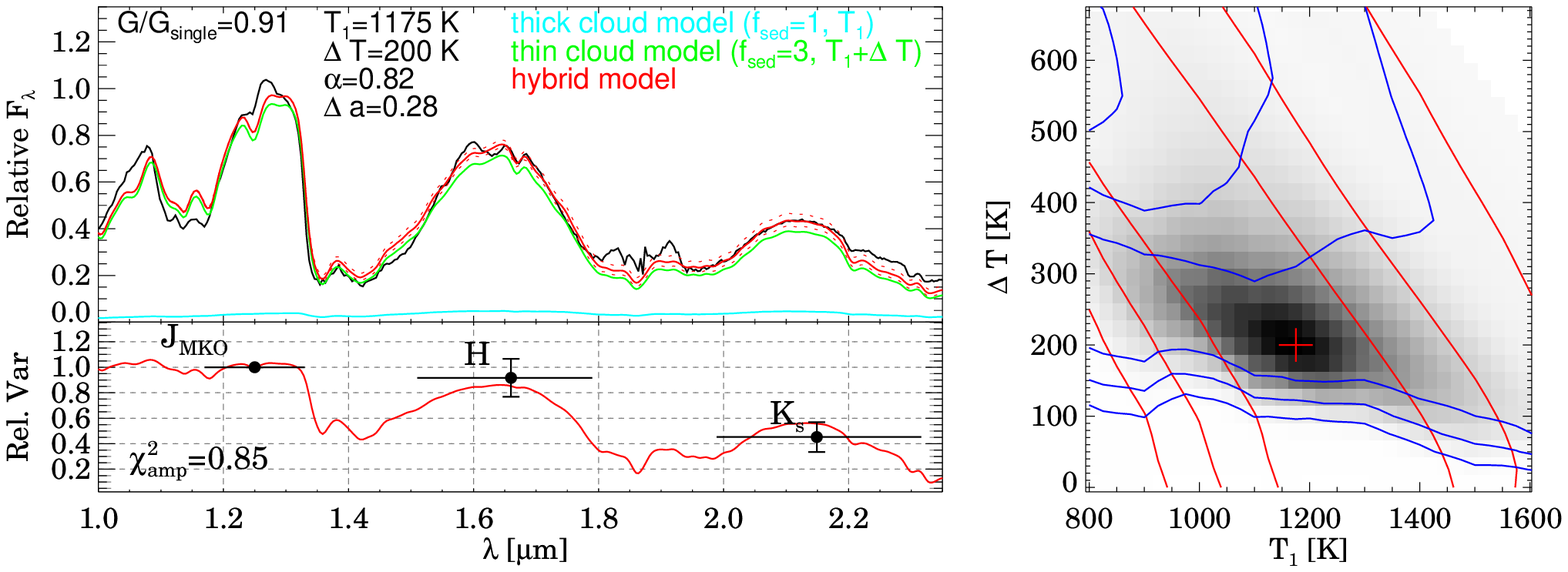} \\[-3mm]
\includegraphics[width=5.5in]{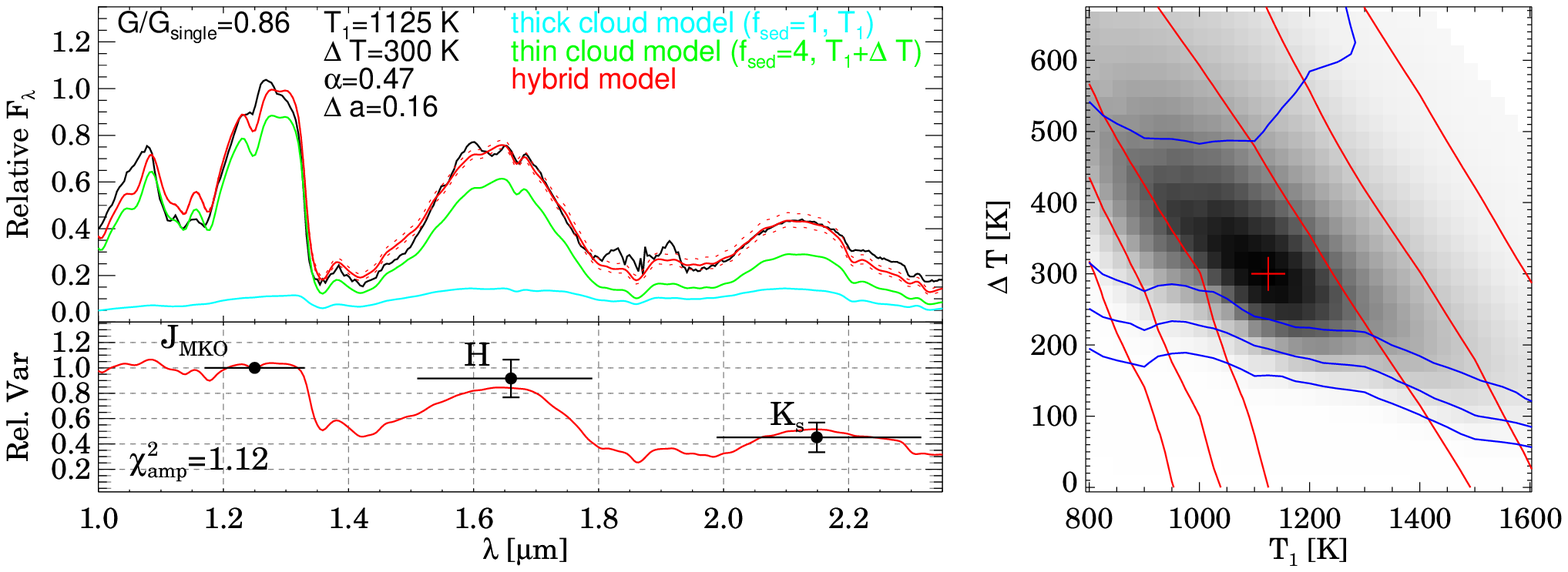} \\[-3mm]
\includegraphics[width=5.5in]{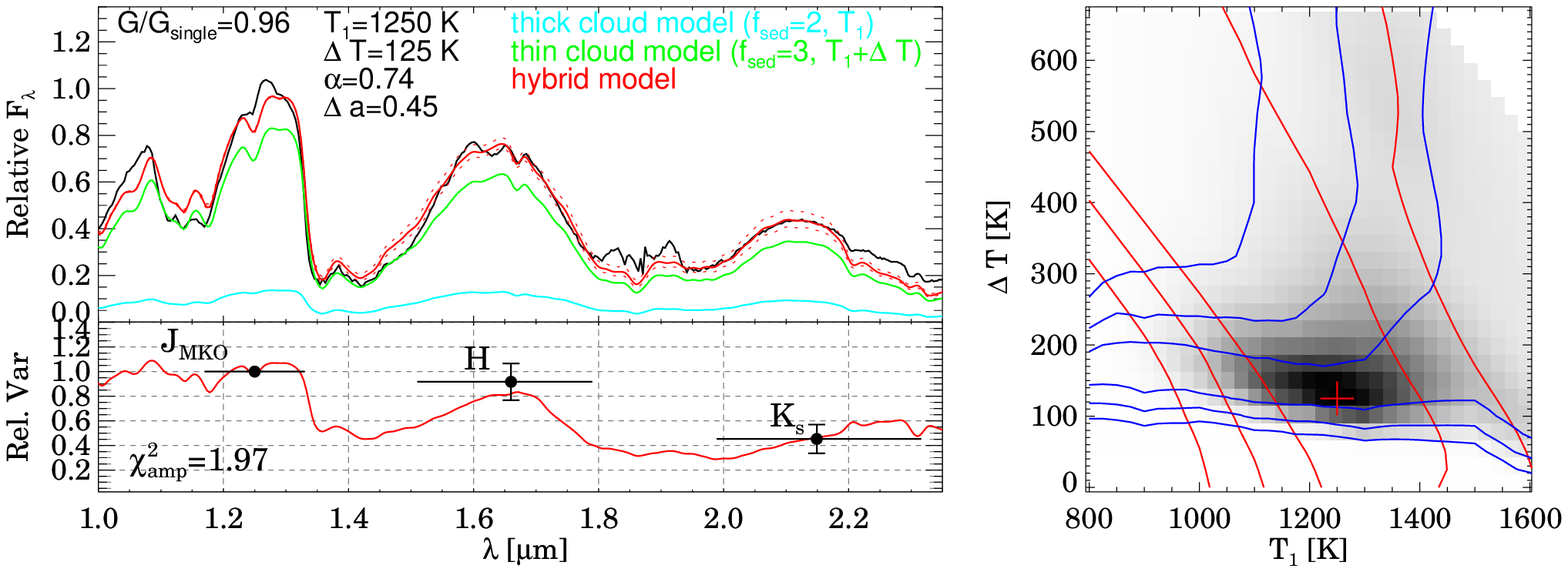} \\[-3mm]
\includegraphics[width=5.5in]{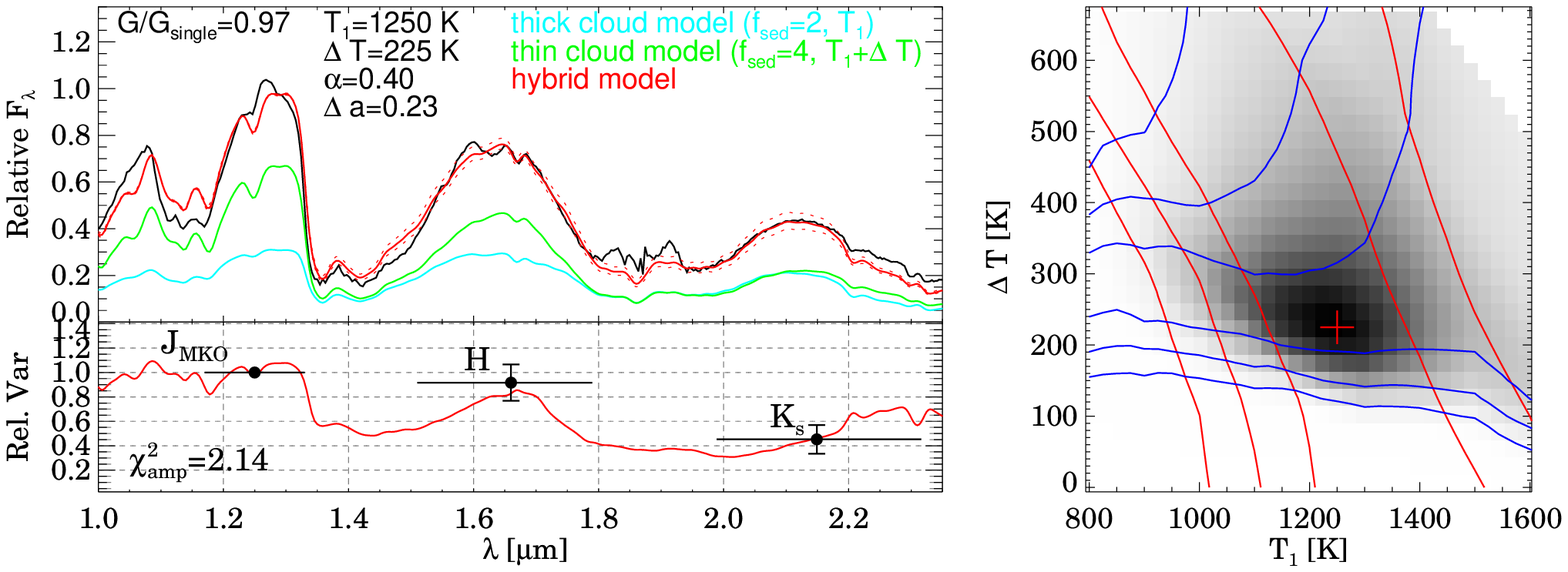}
\vspace{0mm}
\caption{Same as figure \ref{fig:contours}, but with amplitude ratios constrained using data from the 01 Oct 2009 epoch.}
\label{fig:extra_mod2}
\end{figure}

\colb{In section \ref{sect:mod} we presented simultaneous model fits to both the NIR spectrum and multi-color variability of 2M2139.  The model fits presented were constrained using $A_H/A_J$ and $A_{K_s}/A_J$ amplitude ratios measured from the simultaneous $JHK_s$ light curves obtained on 30 Sep 2009 (figure \ref{fig:contours}).   
Here we provide equivalent model fits constrained using amplitude ratios measured from the other two epochs for which we obtained simultaneous $JHK_s$ light curves:  26 Sep 2009 (figure \ref{fig:extra_mod1}), and 01 Oct 2009 (figure \ref{fig:extra_mod2}).}

\clearpage

\bibliographystyle{/home/jackie/manuscripts/astronat/apj/apj}
\bibliography{/home/jackie/manuscripts/lib}

\end{document}